\documentclass{nature}
\def\degree{\hbox{$^\circ$}}
\usepackage[pdftex]{graphicx}
\usepackage{textpos}
\usepackage{lineno}
\usepackage{enumitem}
\usepackage{hyperref}
\usepackage{amsmath}
\usepackage{verbatim} 

\title{Dynamic Europa ocean shows transient Taylor columns and convection driven by ice melting and salinity}

\author{Yosef Ashkenazy$^{1\ast}$\& Eli Tziperman$^{2}$}

\begin{document}
\raggedright


\maketitle

\begin{affiliations}
 \item Department of Solar Energy and Environmental Physics, The Blaustein Institutes for Desert Research, Ben-Gurion University of the Negev, Midreshet Ben-Gurion, 84990, Israel.
 \item Department of Earth and Planetary Sciences and School of Engineering and Applied Sciences, Harvard University, 20 Oxford Street, Cambridge, Massachusetts 02138, USA.
\item[$^\ast$] To whom correspondence should be addressed (ashkena@bgu.ac.il).
\end{affiliations}

\date{}

\newpage

\begin{abstract}
  The deep ($\sim$100 km) ocean of Europa, Jupiter's moon, covered by a thick icy shell, is one of the most probable places in the solar system to find extraterrestrial life. Yet, its ocean dynamics and its interaction with the ice cover have received little attention. Previous studies suggested that Europa's ocean is turbulent using a global model and taking into account non-hydrostatic effects and the full Coriolis force. Here we add critical elements, including consistent top and bottom heating boundary conditions and the effects of icy shell melting and freezing on ocean salinity. We find weak stratification that is dominated by salinity variations. The ocean exhibits strong transient convection, eddies, and zonal jets. Transient motions organize in Taylor columns parallel to Europa's axis of rotation, are static outside of the tangent cylinder and propagate equatorward within the cylinder.   The meridional oceanic heat transport is intense enough to result in a nearly uniform ice thickness, that is expected to be observable in future missions.
\end{abstract}

\section{Introduction}

The possibility of life outside Earth has long-fascinated humankind, and Europa, one of the four Galilean moons of Jupiter, is often mentioned as a candidate\cite{Chyba-Phillips-2001:possible, Hand-Chyba-Priscu-et-al-2009:astrobiology,Pappalardo-Vance-Bagenal-et-al-2013:science} due to its deep ($\sim$100 km) ocean\cite{Cassen-Reynolds-Peale-1979:there, Carr-Belton-Chapman-et-al-1998:evidence, Kivelson-Khurana-Russell-et-al-2000:galileo} that underlies a thick icy shell (several to tens of km)\cite{Billings-Kattenhorn-2005:great,Cassen-Reynolds-Peale-1979:there, Carr-Belton-Chapman-et-al-1998:evidence, Hussmann-Spohn-Wieczerkowski-2002:thermal, Tobie-Choblet-Sotin-2003:tidally}). Europa has a relatively young surface\cite{Bierhaus-Zahnle-Chapman-et-al-2009:europa}, indicating active ice shell tectonics\cite{Carr-Belton-Chapman-et-al-1998:evidence}, and exhibiting chaotic terrain patterns\cite{Cassen-Reynolds-Peale-1979:there, Schmidt-Blankenship-Patterson-et-al-2011:active}. The existence of an ocean under the icy shell is indicated by the observed induced magnetic field\cite{Khurana-Kivelson-Stevenson-et-al-1998:induced}, the indications of ice tectonics\cite{Pappalardo-Belton-Breneman-et-al-1999:does} and perhaps also by water vapor plumes over Europa's mid-southern latitudes\cite{Roth-Saur-Retherford-et-al-2014:transient, Sparks-Hand-McGrath-et-al-2016:probing}.

Europa's ocean dynamics have been studied using a variety of models and mechanisms\cite{Thomson-Delaney-2001:evidence, Goodman-Collins-Marshall-et-al-2004:hydrothermal, Melosh-Ekholm-Showman-et-al-2004:temperature, Tyler-2008:strong, Vance-Goodman-2009:oceanography, Goodman-2012:tilted, Goodman-Lenferink-2012:numerical, Soderlund-Schmidt-Wicht-et-al-2014:ocean,Gissinger-Petitdemange-2019:magnetically}. It has been suggested that localized ocean convection plumes may underlie the observed surface patterns of Europa\cite{Thomson-Delaney-2001:evidence, Goodman-Collins-Marshall-et-al-2004:hydrothermal, Goodman-Lenferink-2012:numerical}. On Earth, due to the very low oceanic aspect ratio (depth over horizontal scale, $\sim 10^{-3}$), only the vertical component of the Coriolis force is relevant. However, the aspect ratio of Europa's ocean is much higher ($\sim 1/16$), and thus the horizontal components of the Coriolis force must be included and have been suggested to result in convection plumes that are parallel to the axis of rotation\cite{Vance-Goodman-2009:oceanography, Goodman-2012:tilted, Soderlund-Schmidt-Wicht-et-al-2014:ocean, Soderlund-2019:ocean}. Scaling arguments were used to suggest the existence of alternating zonal jets\cite{Vance-Goodman-2009:oceanography}, and tidal forcing was proposed to lead to Rossby-Haurwitz waves and thus to oceanic tidal dissipation\cite{Tyler-2008:strong}. Tides can also excite internal waves\cite{Rovira-Navarro-Rieutord-Gerkema-et-al-2019:do} and libration-driven elliptical instability can also drives ocean motions\cite{Lemasquerier-Grannan-Vidal-et-al-2017:libration}. A recent study of Europa's ocean\cite{Soderlund-Schmidt-Wicht-et-al-2014:ocean, Soderlund-2019:ocean} used a global model, taking into account elements such as non-hydrostatic effects and the full Coriolis force, to study the ocean dynamics, and reported a wide low-latitude eastward jet, a high-latitude westward jet, and a rich eddy field. However, the model was adopted from core convection applications and therefore neglected salinity and ice freezing and melting effects that are shown below to dominate those of temperature; it also used upper and lower boundary conditions of prescribed temperature.

Here we show that a more self-consistent formulation, of prescribed bottom heat flux, and a top boundary condition that represents the full interaction with the icy shell and the resulting heat and fresh water fluxes, lead to a very different ocean temperature distribution. Our resolution is higher than that used previously by an order of magnitude, and the viscosity accordingly lower, allowing interesting small scale features to appear.

\section{Results}

\subsection{The model.}

We use a very high-resolution ocean General Circulation Model (GCM), the MITgcm\cite{Marshall-Adcroft-Hill-et-al-1997:finite,MITgcm-manual-github:mitgcm} to investigate the ocean dynamics of Europa, first in a 2d (latitude-depth) configuration, and then in a near pole-to-pole 3d geometry. While the 2d simulations lack several important physical processes, these simulations provide invaluable insight into several critical elements that cannot be addressed in 3d, mostly due to computational cost. We include all components of the Coriolis force, and use the full, non-hydrostatic dynamics. We use a prescribed heat flux as a boundary condition at the bottom rather than prescribing the temperature. This allows the temperature, and in particular the vertical temperature gradient (stratification) to be determined by the model. We follow the modern oceanographic literature and use a three equation formulation\cite{Losch-2008:modeling} (Methods, subsection 3-equation top boundary condition formulation) of the interaction between the icy shell and the ocean temperature and salinity fields, which takes into account the effects of freezing and melting of the icy shell, and diffusion of heat through the ice, on the temperature and salinity. The icy shell is assumed of uniform thickness, an assumption that we show below to be self-consistent with the calculated ocean meridional heat fluxes that were shown previously\cite{Ashkenazy-Sayag-Tziperman-2018:dynamics} to lead to a uniform ice thickness. Estimates of the mean salinity of Europa's ocean vary widely\cite{Hand-Chyba-2007:empirical}, and we choose a value that is close to the lower end of estimates, of 50 ppt (g/kg). We later analyze the sensitivity to this choice.

\subsection{2d model results: Stratification, salinity and Taylor columns.}

The 2d (latitude-depth) simulations (Fig.~\ref{fig:fig-2d-results-tracers}) show that the bottom geothermal heating results in a (potential) temperature at depth that is higher than near the ice-ocean interface by a mere 0.01\degree{C} (Fig.~\ref{fig:fig-2d-results-tracers}a), suggesting that the ocean is well mixed. In addition, note several interesting features. First, surprisingly, the coldest water is at low latitudes, despite the much warmer low-latitude ice surface temperature\cite{Ojakangas-Stevenson-1989:thermal, Ashkenazy-2019:surface}, in contradiction to the findings of previous studies of Europa's ocean\cite{Soderlund-Schmidt-Wicht-et-al-2014:ocean,Soderlund-2019:ocean}. This is explained below as an effect of the Taylor columns discussed there. The ocean is stably stratified at high latitudes and unstably at low latitudes (Fig.~\ref{fig:fig-2d-results-tracers}c), as opposed to the globally unstable stratification imposed in the above mentioned previous studies. Water density variations are dominated by salinity variations, which dwarf the effects of temperature variations ($\beta\Delta S/\alpha\Delta T\gg 1$, where $\alpha$ and $\beta$ are the temperature and salinity expansion coefficients). Previous studies suggest that the salinity may in fact be even higher than assumed here\cite{Hand-Chyba-2007:empirical}. In that case, the salinity gradients due to melting and freezing are expected to be even larger, as salinity rate of change is proportional to the fresh water forcing times the mean salinity (e.g., in the limit of a fresh ocean evaporation does not lead to salinity changes). We therefore focus on the sensitivity of our results to lower mean salinity values and show below that for a wide range of parameters the idea that salinity dominates density variations is robust (Supplementary Figs.~1-3). The zonal velocity (Fig.~\ref{fig:fig-2d-results-vel}a,d) is westward in the low-latitude upper ocean and eastward elsewhere, with a typical velocity of a few cm per second. The deep equatorial zonal velocity is positive (eastward), indicating a superrotation, further discussed below.

Prominent arc-like structures appear in all fields (Figs.~\ref{fig:fig-2d-results-tracers},\ref{fig:fig-2d-results-vel}), which reflect features parallel to the rotation axis when plotted in spherical geometry (Fig.~\ref{fig:fig-2d-results-vel}e). These are Taylor columns with ocean velocity nearly independent of the direction parallel to the rotation axis, and expected for an ocean with a nearly uniform density. While these columns were previously anticipated based on scaling arguments\cite{Vance-Goodman-2009:oceanography}, simulated and attributed to convection\cite{Soderlund-2019:ocean}, and seen in simulations of magnetically-driven ocean circulation\cite{Gissinger-Petitdemange-2019:magnetically}, their detailed dynamics, structure, role in setting the large-scale temperature and salinity structure, and their spacing and propagation have not been studied.

The velocity along the Taylor columns fluctuates as one moves from Europa's center outward, between being positive and negative. Accordingly, the heat advection changes sign as well. In the region inside of the tangent cylinder that is aligned with the rotation axis and has the radius of Europa's rocky core, corresponding to latitudes less than about 20\degree\cite{Goodman-2012:tilted}, the columns intersect the ocean bottom and the ice base, and their along-column motions effectively transfer the bottom geothermal heat to the ocean surface. However, within the tangent cylinder, there is no such effective bottom-to-surface heat transport mechanism, as the Taylor columns do not intersect Europa's ocean bottom there, and this results in the colder ocean surface in the equatorial regime seen in Fig.~\ref{fig:fig-2d-results-tracers}a. This leads to freezing there, and thus to brine rejection and to the higher salinity as seen in Fig.~\ref{fig:fig-2d-results-tracers}b.

We find that the spacing between the Taylor columns is less than 20 km (0.75 degree latitude, Fig.~\ref{fig:taylor-spacing}a). In order to analyze the Taylor columns, we project the model's meridional $v$ and vertical $w$ velocity components on the directions parallel and perpendicular to the axis of rotation. The velocity parallel to the axis of rotation, $u_{\rm par}=w\sin\phi+v\cos\phi$ where $\phi$ is the latitude, shows clear Taylor columns in which it is independent of the direction parallel to the axis of rotation (Fig.~\ref{fig:fig-2d-results-vel}b,e), in accordance with the Taylor-Proudman theorem. This parallel velocity is significantly smaller than the zonal velocity (Fig.~\ref{fig:fig-2d-results-vel}a,d) and significantly larger than the velocity perpendicular to the axis of rotation in the latitude-depth plane $u_{\rm per}=w\cos\phi-v\sin\phi$ (Fig.~\ref{fig:fig-2d-results-vel}c). In the zonal momentum budget of the 2d model, the two Coriolis terms dominate the others, so that the momentum balance is $2\Omega w\cos\phi-2\Omega v\sin\phi\approx0$, where $\Omega$ is Europa's rotation rate. This leads to $u_{\rm per}\approx0$, explaining the observation that $u_{\rm per}\ll u_{\rm par}$. The parallel and zonal velocities are symmetric with respect to the equator, while the perpendicular is anti-symmetric, vanishing at the equator.

The distance between the Taylor columns can be estimated using scaling arguments (Methods, subsection The spacing between the Taylor columns). In the zonal and meridional dominant momentum balances, the sum of the two dominant Coriolis terms is balanced by parameterized horizontal viscosity, and one can form a length scale from the two relevant parameters, the horizontal viscosity coefficient $\nu_h$ (m$^2$s$^{-1}$) and the rotation rate $\Omega$ (s$^{-1}$), to find that the relevant length scale is $C\sqrt{\nu_h\sin(\phi)/\Omega}$ where $C\approx14$ is an empirical constant found from the numerical results (Fig.~\ref{fig:taylor-spacing}a). The horizontal viscosity coefficient we used (50 m$^2$s$^{-1}$) represents parameterized viscosity (Methods, subsection Eddy coefficients, subgrid-scale representation). Further verification of the scaling for the distance between columns is obtained below in the 3d runs, where the effective resolved eddy viscosity is found to be larger (300 m$^2$s$^{-1}$, see Methods, subsection Estimating eddy coefficients), and the column distance is indeed larger (Supplementary Fig.~6). While this particular spacing is likely sensitive to model assumptions, the qualitative dynamical insights obtained should be valid. Additional simulations suggest that the distance between the Taylor columns in the high latitudes scales like the square root of the ocean depth. The existence of Taylor columns in Europa, and the corresponding zonal jet structure was predicted by ref. \cite{Vance-Goodman-2009:oceanography} to be related to the Rhines scale, although our findings regarding the spacing between the columns is different from the Rhines scale scaling predicted in that work. These Columns seem to also appear in one of the simulations of\cite{Soderlund-2019:ocean} that was characterized by low viscosity, although no detailed analysis was provided.

In absence of dissipation the Taylor columns were predicted to be at a fixed latitude, as the potential vorticity ($q=(2\Omega+\zeta)/h$, where $h$ is the column height, which depends on latitude) is preserved\cite{Vance-Goodman-2009:oceanography}. Yet we find the columns to show prominent equatorward propagation outside of the tangent cylinder (Fig.~\ref{fig:taylor-spacing}b,c) whose mechanism would require further elucidation in future work. No propagation is visible within the tangent cylinder (Fig.~\ref{fig:taylor-spacing}b,c). We also note the oscillatory variations along the maximum/minimum lines (Fig.~\ref{fig:taylor-spacing}b,c). While the discussion in this subsection clearly explains the structure and spacing of the (2d) Taylor columns, the necessarily-over simplified eddy viscosity formulation used may affect the simulation. Below we show, based on 3d simulations, that in fact the eddy parameterized coefficient is much larger than the one used in the 2d results, lending credibility to the 2d results.

\subsection{3d model results: eddies, convecting plumes.}

We next consider a 3d simulation of Europa's ocean at very high resolution (1/24 of a degree, compared with $\sim$1\degree{} of previous studies\cite{Soderlund-Schmidt-Wicht-et-al-2014:ocean}). The model spans 30 degree longitude and we assume periodic boundary conditions in the zonal direction. The added 3rd, zonal, dimension allows for waves and eddies to develop and enables us to examine the interaction of eddies with the Taylor columns and convection (Figs.~\ref{fig:3d-results-temp},\ref{fig:3d-results-vars}). The Taylor columns now appear most prominently in the simulated meridional and vertical velocity fields as isolated columns with a width and separation of about 20--50 km (Supplementary Figs.~7-11
). The columns are again largely aligned with the rotation axis\cite{Vance-Goodman-2009:oceanography, Soderlund-2019:ocean} as in the 2d model. At low latitudes, high-salinity downward convection plumes originate from the ice-ocean interface (Fig.~\ref{fig:3d-results-temp}b) due to brine-rejection during the freezing process that was not included in previous studies of Europa's ocean (upper part of Fig.~\ref{fig:3d-results-temp}b, and Supplementary Videos 1 and 2). Upward plumes are seen in Fig.~\ref{fig:3d-results-temp}c to originate from the ocean bottom due to the geothermal heating there. These convective plumes are nearly perpendicular to the Taylor columns near the equator, in contradiction to expectations based on regional simulations\cite{Vance-Goodman-2009:oceanography, Goodman-2012:tilted, Goodman-Lenferink-2012:numerical}, and are also visibly advected by the mean zonal flows. The orthogonality of the low-latitude convection and Taylor columns suggests that these two classes of motion are distinct. Furthermore, the 2d sensitivity run shown in Supplementary Figs.~2-5 with a vanishing mean salinity shows a regime that is completely stratified with no convection occurring anywhere, yet with Taylor columns prominent in the zonal and meridional velocity components. The energy source for the Taylor columns, and in particular what is the specific instability mechanism involved, requires further study.

The temperature field is clearly turbulent, showing richly complex eddy filaments (Fig.~\ref{fig:3d-results-temp}a, see supplementary animations). The above relation between viscosity and Taylor column spacing, together with the fact that the Taylor column spacing is larger in the 3d simulation, suggests that the effective eddy viscosity due to resolved eddies is about 15 times larger than the small explicit parameterized viscosity used in the 3d runs for numerical stability, following common ocean modeling practice. This is consistent with an explicit estimate of the eddy coefficients calculated from the 3d runs (Supplementary Fig.~12). As a result of the eddies and convection plumes, the Taylor columns are less persistent along the direction parallel to the rotation axis than in the 2d simulations (Figs.~\ref{fig:3d-results-temp}, \ref{fig:3d-results-vars}, and Supplementary Figs. 8-11
). The existence of waves and eddies in the 3d simulation also affects mean flows. The zonal velocity is typically several cm~s$^{-1}$ (Fig.~\ref{fig:3d-results-vars}b), 1-2 orders of magnitude smaller than the previously reported velocities\cite{Soderlund-Schmidt-Wicht-et-al-2014:ocean,Soderlund-2019:ocean}; thus, our estimate for Europa's ocean kinetic energy (see below) is several orders magnitude smaller than that of these previous studies. Note in particular differences in the vertical structure of the zonal jets along the equator, between the 2d (Fig.~\ref{fig:fig-2d-results-vel}a,d) and 3d (Fig.~\ref{fig:3d-results-vars}b) simulations; see also Supplementary Fig.~13. The 2d zonal flow shows superrotation only at depth, while the 3d ones shows it at all depths. The 2d vertical shear with superrotation at depth and a retrograde surface current is likely supported by an eddy flux of zonal momentum toward the rotation axis. The 3d superrotation can be driven by Rossby waves that are possible in this configuration. Moreover, the increasing 2d equatorial zonal velocity with depth is consistent with the thermal-wind relation (i.e., $u_z=\frac{g}{2\Omega a \rho_0\sin(\phi)} {\rho_\phi}$). In the 3d case, we find an additional term that cannot be neglected (i.e., $a\,{\sin(\phi)} u_z +(\cos(\phi) u)_\phi=\frac{g}{2\Omega \rho_0} {\rho_\phi}$), explaining the different vertical structure of the zonal velocity $u$. The zonal domain of the model spans only 30\degree{} for computational efficiency. However, because the typical size of eddies in the 3d simulation is much smaller than the zonal extent, we expect the eddy dynamics to be similar had we considered a fuller zonal extent of 360\degree{}.

\begin{figure}
  \centerline{\includegraphics[width=20pc]{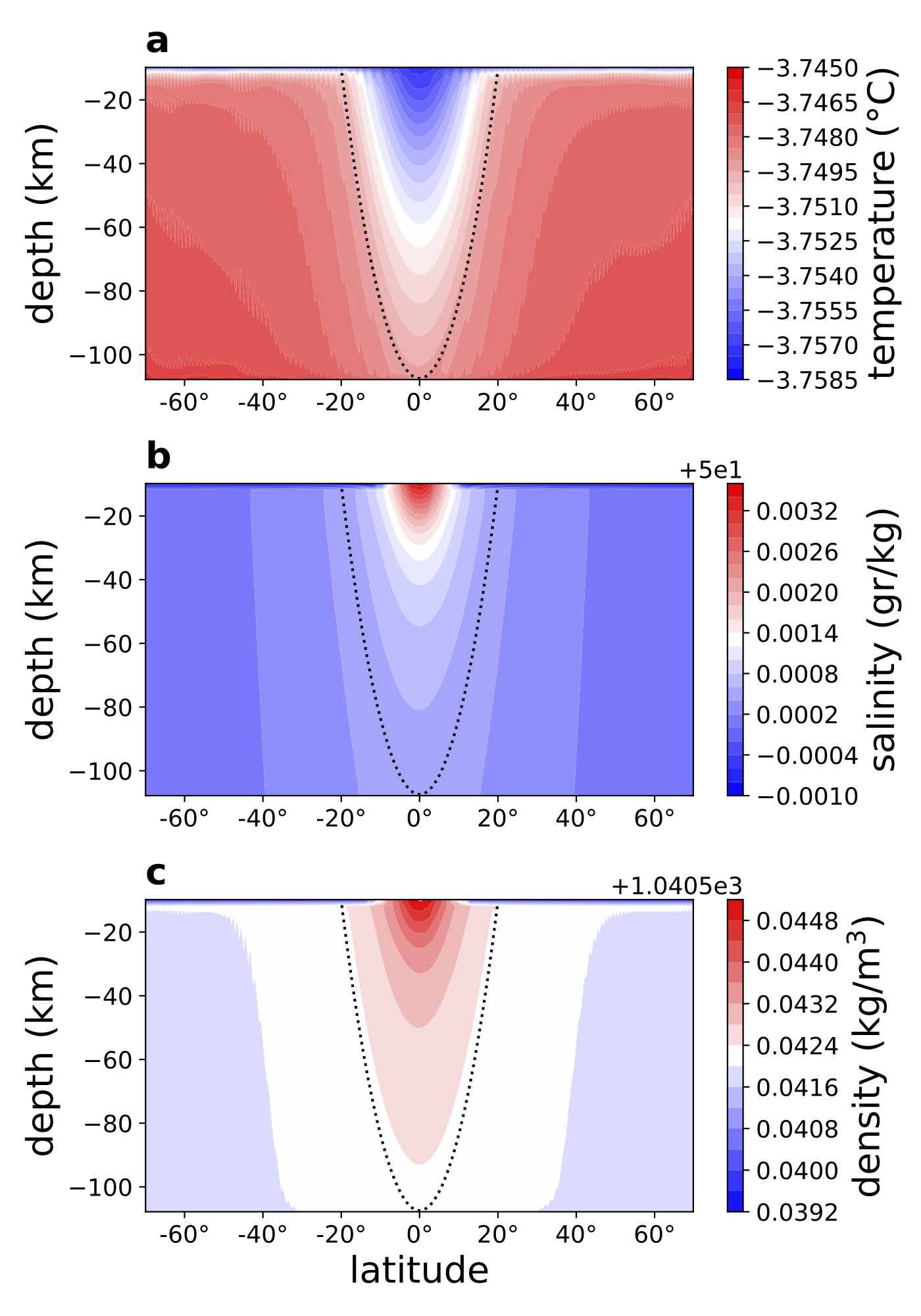}}
  \caption{{\bf 2d simulation results--tracers.}  Latitude-depth snapshot plots of {\bf a} temperature (\degree{C}), {\bf b} salinity (gr kg$^{-1}$), and {\bf c} density (kg m$^{-3}$). The dotted lines in these panels shows the tangent cylinder. 
    \label{fig:fig-2d-results-tracers}}
\end{figure}

\begin{figure}
  \begin{center}
    \includegraphics[width=30pc]{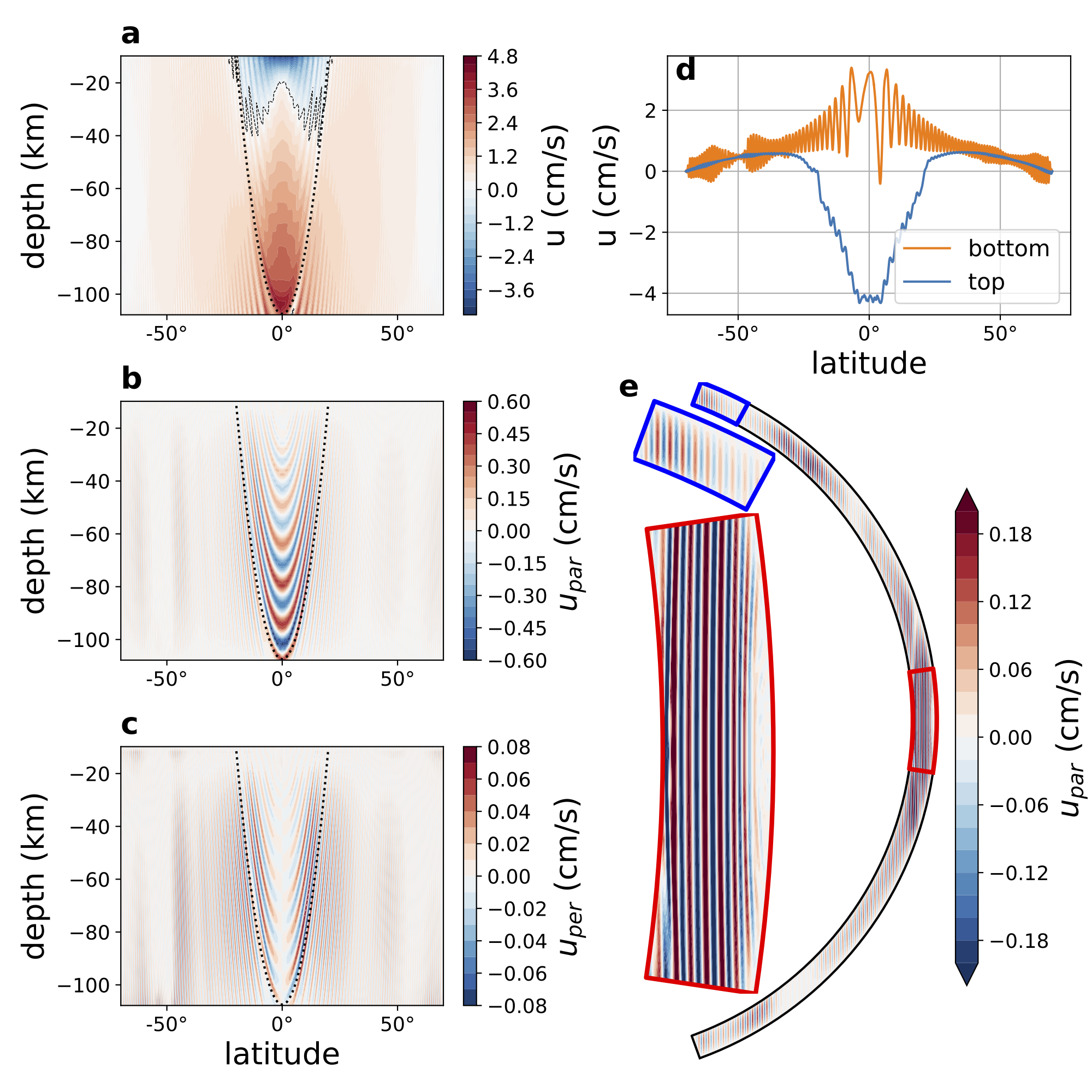}
    \end{center}
  \caption{{\bf 2d simulation results--velocity.}  Latitude-depth snapshot plots of {\bf a} zonal velocity (cm s$^{-1}$, dashed line shows the zero contour), {\bf b} velocity parallel to the axis of rotation, $u_{\rm par}$ (cm s$^{-1}$), and {\bf c} velocity perpendicular to the axis of rotation, $u_{\rm per}$ (cm s$^{-1}$).  The dotted lines in these panels shows the tangent cylinder. {\bf d} Zonal velocity at the top and bottom of the ocean (cm s$^{-1}$) as a function of latitude.  {\bf e} The spherical presentation of the velocity parallel to the axis of rotation $u_{\rm par}$ (also shown in panel {\bf b}), demonstrating Taylor columns that are parallel to the axis of rotation. The ocean depth extent is 100 km in all three frames shown, where the latitudinal extent of the main (black) frame is 70\degree{S}-70\degree{N}, the low latitude (red) frame is 8\degree{S}-8\degree{N}, and the high latitude (blue) frame is 61\degree{N}-70\degree{N}.
    \label{fig:fig-2d-results-vel}}
\end{figure}

While the stratification is very weak and the ocean well-mixed (Fig.~\ref{fig:fig-2d-results-tracers}a--c), as in the 2d case, the extent of unstable water column with heavy water above light water is more limited in the 3d case (compare Fig.~\ref{fig:fig-2d-results-tracers}a--c and Figs.~\ref{fig:3d-results-temp}c, \ref{fig:3d-results-vars}a). This is because the eddies in the 3d simulation strengthen the stratification, by converting potential energy into kinetic energy, as was suggested to be the case for Earth's snowball ocean\cite{Jansen-2016:turbulent,Ashkenazy-Tziperman-2016:variability}. The characteristic time of convection may be estimated via the buoyancy frequency, $N^2=-g/\rho_0\partial \rho/\partial z\approx (g/H)(\Delta\rho/\rho_0)$ where negative/positive $N^2$ indicates statically unstable/stable water column. We find typical buoyancy and convection time scales, corresponding to positive and negative values of the buoyancy frequency, with corresponding time scales $2\pi/|N|$ that both exceed 50 days. The positive values represent time scales of internal waves in Europa's ocean and are much larger than the corresponding time scales on Earth, and in fact represent an interesting regime where the Coriolis time scale is shorter than that of buoyancy oscillations. For negative values, the time scales are again much longer than those on Earth, and are consistent with the evolution times seen in the supplementary animations and with previous higher resolution regional runs\cite{Goodman-Lenferink-2012:numerical}.

Refs. \cite{Vance-Brown-2005:layering,Vance-Goodman-2009:oceanography} suggested the possibility of double diffusion in Europa's ocean. We find in the low latitudes in the 3d model and in the high latitudes of the 2d model, warm salty water under a surface layer of 1--2 grid points that is cold and fresh, and where the stratification is stable. While this is a scenario that can, in principle, lead to double diffusion and therefore to an enhanced vertical mixing, the surface layer is hardly resolved numerically and our results therefore do not seem to provide definite evidence for or against the idea that double diffusion may play a role in Europa's ocean.

One can get further insight into the eddy field from an energetic point of view. The oceanic available potential energy (APE) is the potential energy that may be converted into kinetic energy (KE). The ratio between the APE and the KE provides a measure of the efficiency of kinetic energy extraction from the stratification, and an indication of the source of eddy kinetic energy. For present-day Earth, the ratio between the APE and the KE ocean is over 33,000\cite{Vonstorch-Eden-Fast-et-al-2012:estimate}. Europa's APE calculated following\cite{Huang-1999:mixing} is $2.3\times 10^{18}$J for our 3d model domain, only 190 times larger than the KE, a factor similar to that of the Snowball Earth ocean, where APE/KE was estimated at about 300\cite{Ashkenazy-Tziperman-2016:variability}. While the estimate of\cite{Vonstorch-Eden-Fast-et-al-2012:estimate} may be sensitive to the mean vertical density gradient used, we use the method of \cite{Huang-1999:mixing} because it does not depend on this gradient and is more appropriate for a very weakly stratified ocean. The ratio being much smaller in these Europa simulations is therefore a robust result. On Earth, macro-turbulence in the ocean and atmosphere is generated mostly by the drawing down of APE via baroclinic instability. Our results suggest that in Europa's ocean, convective plumes and barotropic instability of the zonal jets play a more prominent role in the generation of ocean macro-turbulence relative to baroclinic instability.

A previous study of the dynamics of the icy shell\cite{Ashkenazy-Sayag-Tziperman-2018:dynamics} showed that an efficient meridional ocean heat transport can lead to a uniform shell thickness. The geothermal heat flux entering the ocean from below is larger than the heat escaping through the ice in the tropics and smaller at high latitudes, due to the meridional ice surface temperature gradient\cite{Ojakangas-Stevenson-1989:thermal,Ashkenazy-2019:surface}. This would lead to melting at low latitudes and freezing at high latitudes, leading to ice thickness gradients balanced by ice flow\cite{Ashkenazy-Sayag-Tziperman-2018:dynamics}. However, an efficient poleward ocean heat transport can carry the excess heat meridionally, and thus overcome the tendency toward meridional ice thickness gradients, and result in almost uniform ice thickness (Fig.~\ref{fig:fig4}b). The meridional heat transports of the 2d and 3d ocean simulations are shown by the solid lines in Fig.~\ref{fig:fig4} to be poleward, and have maximum values of about $0.5\times 10^{11}$W and $1.5\times 10^{11}$W, correspondingly. These estimates of the meridional heat fluxes in a full ocean model are at least four times larger than the heat transport estimated by\cite{Ashkenazy-Sayag-Tziperman-2018:dynamics}, because they include the contribution due to latent heat of freezing at the equator and melting at the poles, not included in previous studies.

The meridional heat flux without the latent heat contribution is shown by the green solid curve. This heat flux is determined, as explained above, by the geothermal and surface heat fluxes calculated for the assumed uniform thickness ice shell. The ocean has no difficulty transporting this heat flux in a way that is consistent with the uniform ice shell assumption. An ocean without an efficient meridional heat flux mechanism would have been heated in the tropics and cooled in the high latitude, not being able to reach a steady state. We conclude that the efficient ocean heat transport in our simulation is self-consistent with the assumption of a uniform ice thickness, justifying the use of a uniform thickness icy shell in this study. That the 3d meridional heat flux is somewhat larger than the 2d flux is a direct result of the larger latent heat due to freezing in the low latitudes in the 3d case. The difference between the two model configurations is not large, and is within the uncertainty of the internal ocean variability, as estimated for example via the difference between the two hemispheres in the 3d case. Spatial variations in tidal heating within the ice may still cause a range of surface heat fluxes and therefore ice thickness variations\cite{Ojakangas-Stevenson-1989:thermal, Sotin-Tobie-Wahr-et-al-2009:tides}, if the ice is sufficiently thick (thicker than chosen here based on\cite{Chen-Nimmo-Glatzmaier-2014:tidal}) to allow convection.

\section{Discussion}

The 2d and 3d high resolution simulations of Europa's ocean analyzed here show a number of surprising results. While both the temperature and salinity are nearly uniform, salinity gradients, not considered previously, dominate the gradients in ocean water density, and the heaviest water (cold and saline) is found, as a result, at low latitudes. We showed this to be a result of the lack of direct communication along Taylor columns between the ocean bottom and surface in the area outside of the tangent cylinder. Taylor columns that are parallel to the axis of rotation are prevalent and show two regimes\cite{Goodman-2012:tilted}, the low latitudes (outside of the tangent cylinder, equatorward of $\sim$20\degree{}) at which the Taylor columns do not intersect the ocean bottom and extend from one hemisphere to the other, and higher latitudes at which they intersect the ocean bottom. The Taylor columns, which were previously expected not to propagate in latitude due to potential vorticity conservation, and not to occupy the low-latitude regime\cite{Vance-Goodman-2009:oceanography}, exhibit meridional propagation due to frictional effects and occupy the entire ocean. Their spacing was explained above in terms of the rotation rate and viscosity. The 3d simulation shows a rich turbulent eddy flow, as well as convective plumes due to freezing and brine rejection near the ice-ocean interface, and due to geothermal heating from below. The convection plumes are perpendicular to the Taylor columns at low latitudes. We found superrotation at the equator, attributed it to eddy fluxes of zonal momentum and thermal wind balance, and attempted to explain the reasons for the difference in its structure between the 2d and 3d simulations. The meridional heat flux deduced here is much larger than previously estimated\cite{Ashkenazy-Sayag-Tziperman-2018:dynamics}, due to the contribution of the latent heat of freezing that was not considered in previous studies. The ratio between the APE and the KE is significantly smaller than on present-day Earth, yet similar to that estimated for the Snowball Earth ocean\cite{Ashkenazy-Gildor-Losch-et-al-2013:dynamics, Ashkenazy-Tziperman-2016:variability, Jansen-2016:turbulent}.

A few recently submitted manuscripts investigate complementary aspects of the role of salinity in icy satellites to those discussed here, although they do not deal with the eddy motions and Taylor columns analyzed here. Ref. \cite{Kang-Mittal-Bire-et-al-2021:how} examines the effect of ocean salinity on ice thickness and meridional ocean circulation. Ref. \cite{Zeng-Jansen-2021:ocean} examines the effects of low vs high salinity on the circulation and stratification of Enceladus via the suppression of the water density anomaly by ocean salinity (see also our sensitivity experiments, Supplementary Figs.~1-5). Ref. \cite{Lobo-Thompson-Vance-et-al-2020:pole} finds meridional overturning circulation and shallow freshwater polar lenses in Enceladus simulations.

Several of our above predictions may be verified in future missions to Europa, such as the Europa Clipper of NASA\cite{Pappalardo-Prockter-Senske-et-al-2016:science, Howell-Pappalardo-2020:nasas} or JUICE of ESA\cite{Grasset-Dougherty-Coustenis-et-al-2013:jupiter}. These include the uniform icy shell thickness due to the efficient meridional heat transport of Europa's ocean predicted in our simulations. The small meridionally variations of salinity predicted here may similarly be observable in future missions as well through its magnetic signal\cite{Vance-Styczinski-Bills-et-al-2021:magnetic} although this may be challenging. In addition to these observable predictions, the eddy diffusivity and viscosity coefficients -- estimated here from an eddy-resolving simulation of Europa's ocean -- should help in estimating ocean heat generation due to tides\cite{Chen-Nimmo-Glatzmaier-2014:tidal}. Similarly, the weak or even weakly unstable stratification suggests that internal wave breaking may not be a significant factor in tidal dissipation, consistent with previous estimates\cite{Chen-Nimmo-Glatzmaier-2014:tidal}. The libration of the icy shell may be influenced by ocean eddies and ocean currents and this may serve as a way of indirectly observing ocean dynamics. Finally, the study of Europa's ocean may help to better understand the ocean and ice dynamics of other icy moons/planets in the solar system and beyond.

\newpage

\begin{figure}
  \begin{center}
    \includegraphics[width=18pc]{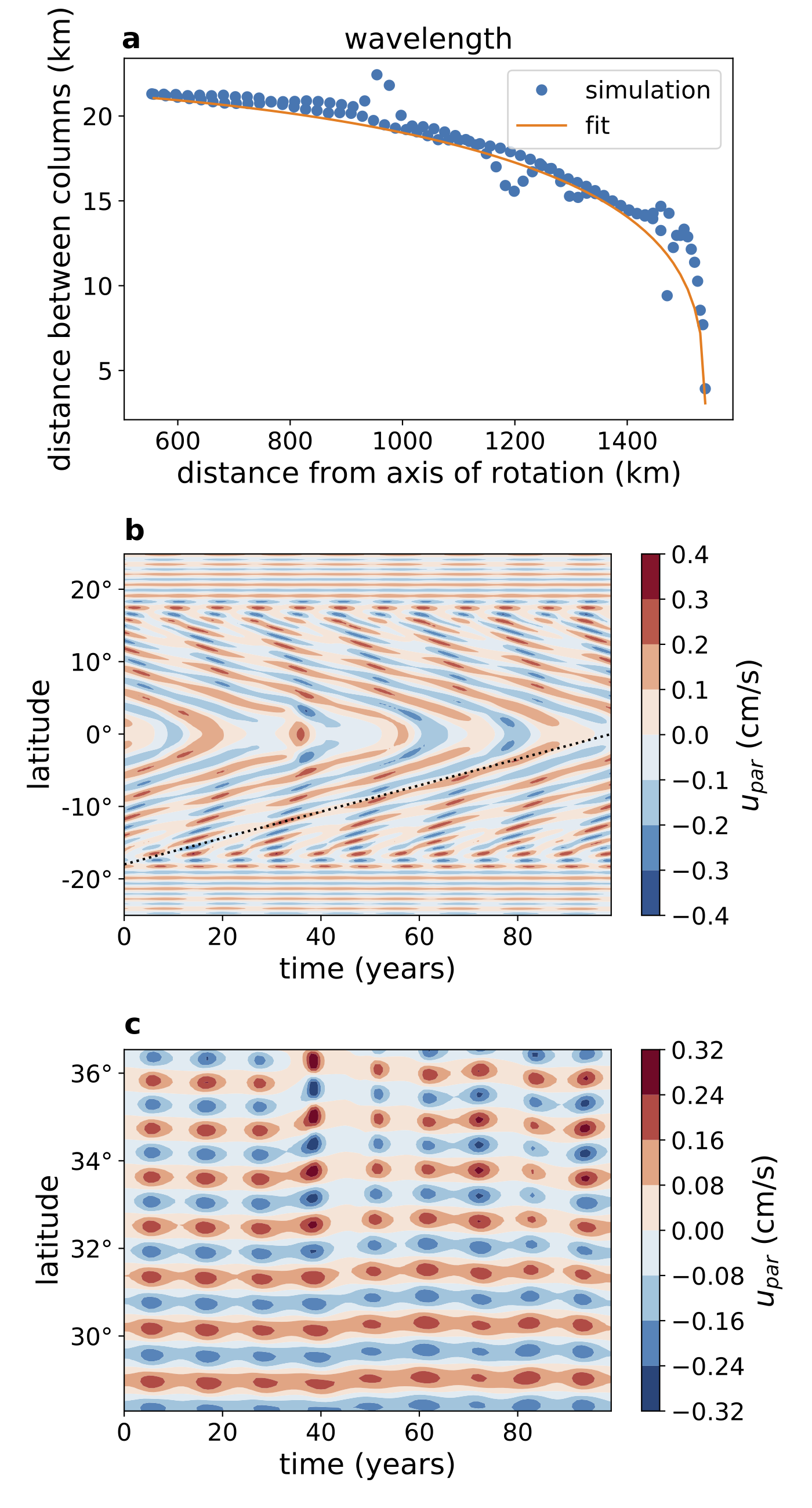}
  \end{center}
  \vskip -1cm
  \caption{{\bf The dynamics and structure of the Taylor columns.} {\bf a} The distance (km) between the Taylor columns in the 2d simulation as a function of the distance from the axis of rotation (km), based on a snapshot of the velocity parallel to the axis of rotation, $u_{\rm par}$ at a depth of 13 km under the ice-ocean interface.  Full circles represent the numerically estimated distances while the solid line represents the functional fit, see text. {\bf b} The velocity parallel to the axis of rotation (cm s$^{-1}$) as a function of latitude and time (Earth years) at a depth of 19 km under the ice. The figure shows the equatorward propagation of the Taylor columns, where the slope of the shown dashed line corresponds to a propagation velocity of 0.18\degree{} per year. {\bf c} Same as \textbf{b}, showing a region inside the tangent cylinder (at depth of 49 km under the ice), where the Taylor columns are static. 
    \label{fig:taylor-spacing}}
\end{figure}

\begin{figure}
  \includegraphics[width=36pc]{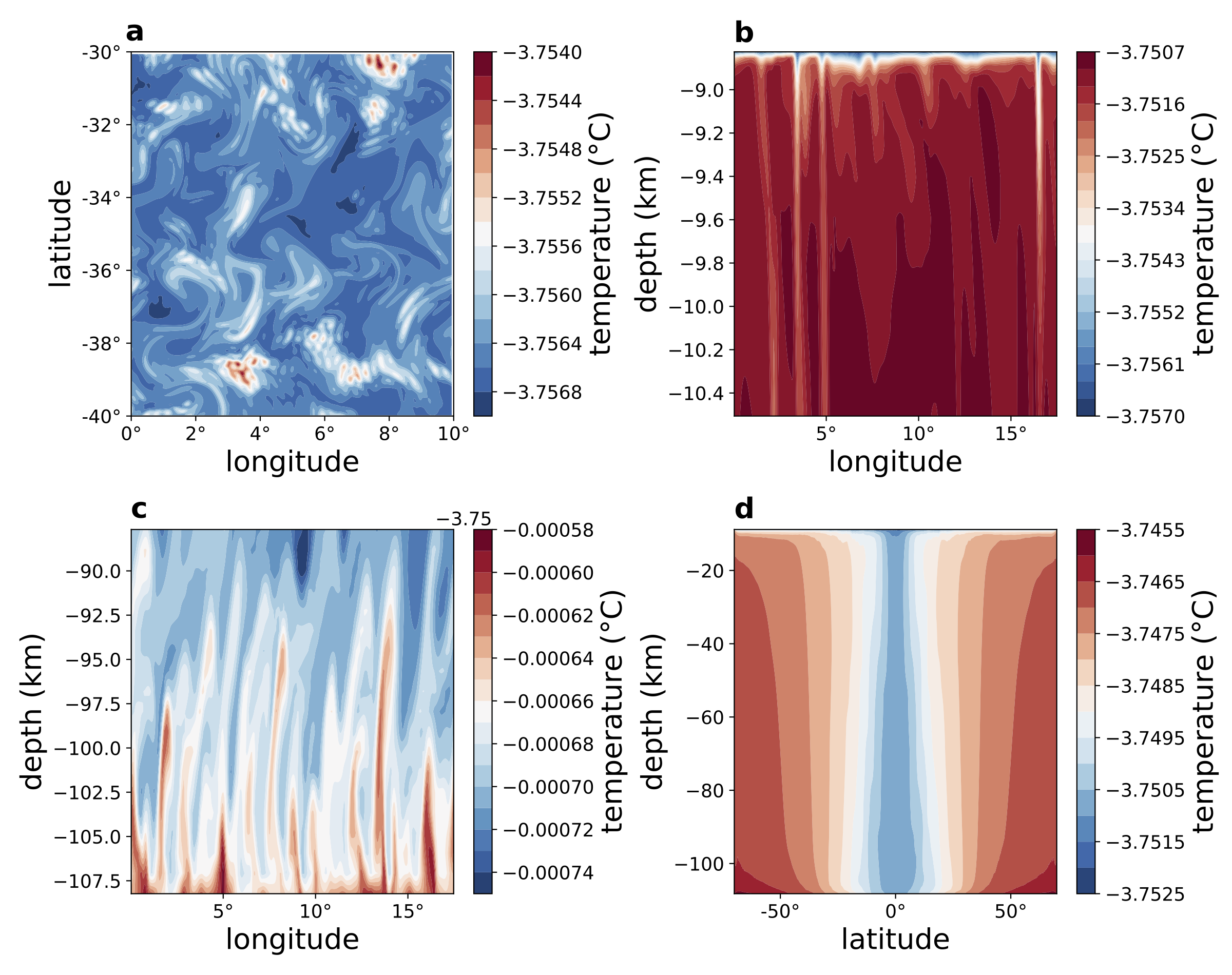}
  \caption{{\bf Results of the 3d simulation--temperature.} {\bf a} Surface ocean temperature (\degree{C}) as a function of longitude and latitude for a $10\times10$ degree region, showing a highly turbulent flow. {\bf b}, {\bf c} Depth-longitude temperature (\degree{C}) sections near the top and bottom of the ocean at the equator, showing downward and upward convection plumes. At the equator, the local vertical (depth) direction is perpendicular to the Taylor columns, and the panels therefore demonstrate that convection does not necessarily occur along the Taylor columns. {\bf d} Latitude-depth plots of zonal mean temperature (\degree{C}).
    \label{fig:3d-results-temp}}
\end{figure}

\begin{figure}
  \centerline{\includegraphics[width=18pc]{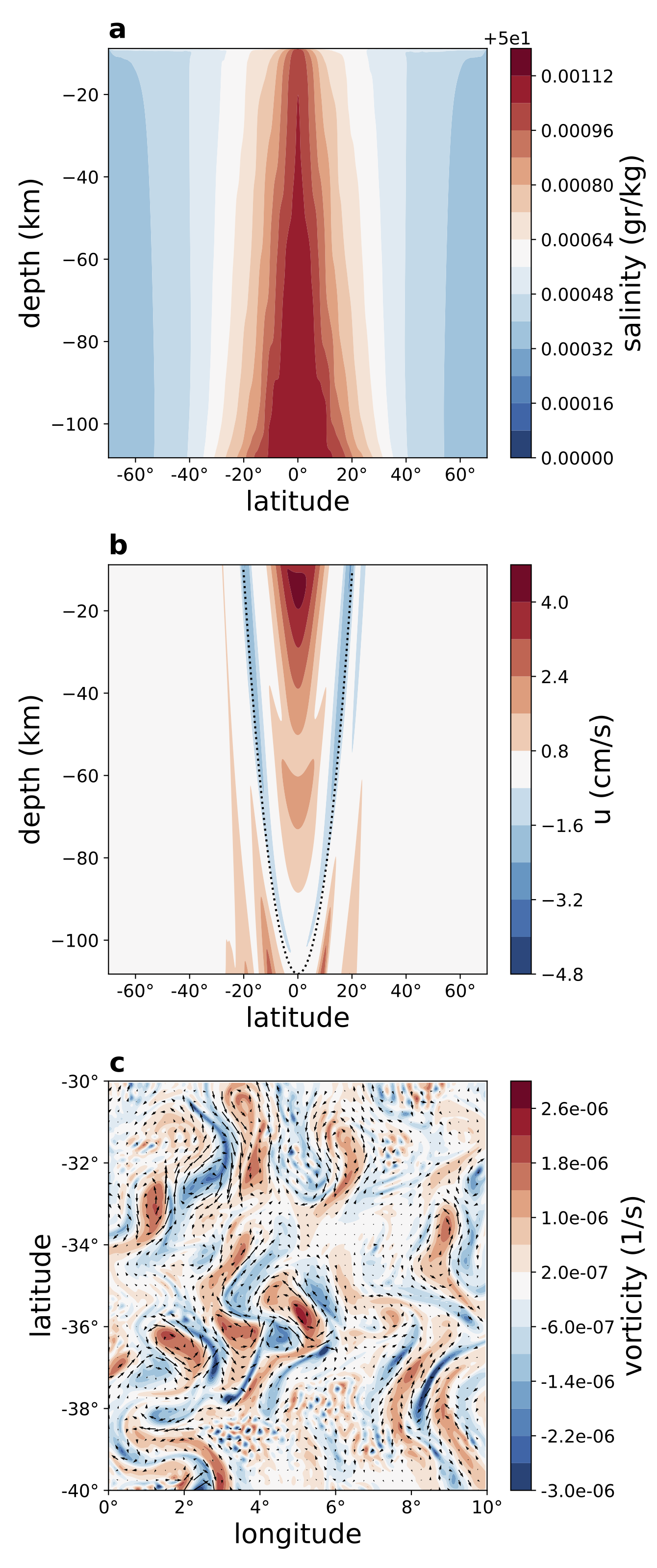}}
  \caption{{\bf Results of the 3d simulation--salinity, zonal velocity, and vorticity.} {\bf a}, {\bf b} Latitude-depth plots of zonal mean salinity (gr/kg)  and zonal velocity (cm s$^{-1}$). {\bf c} Vorticity (s$^{-1}$) and velocity field (arrows) at the ocean surface, as a function of longitude and latitude, demonstrating the rotational fluid velocity around the Taylor columns.
    \label{fig:3d-results-vars}}
\end{figure}

\begin{figure}
\centerline{\includegraphics[width=0.95\linewidth]{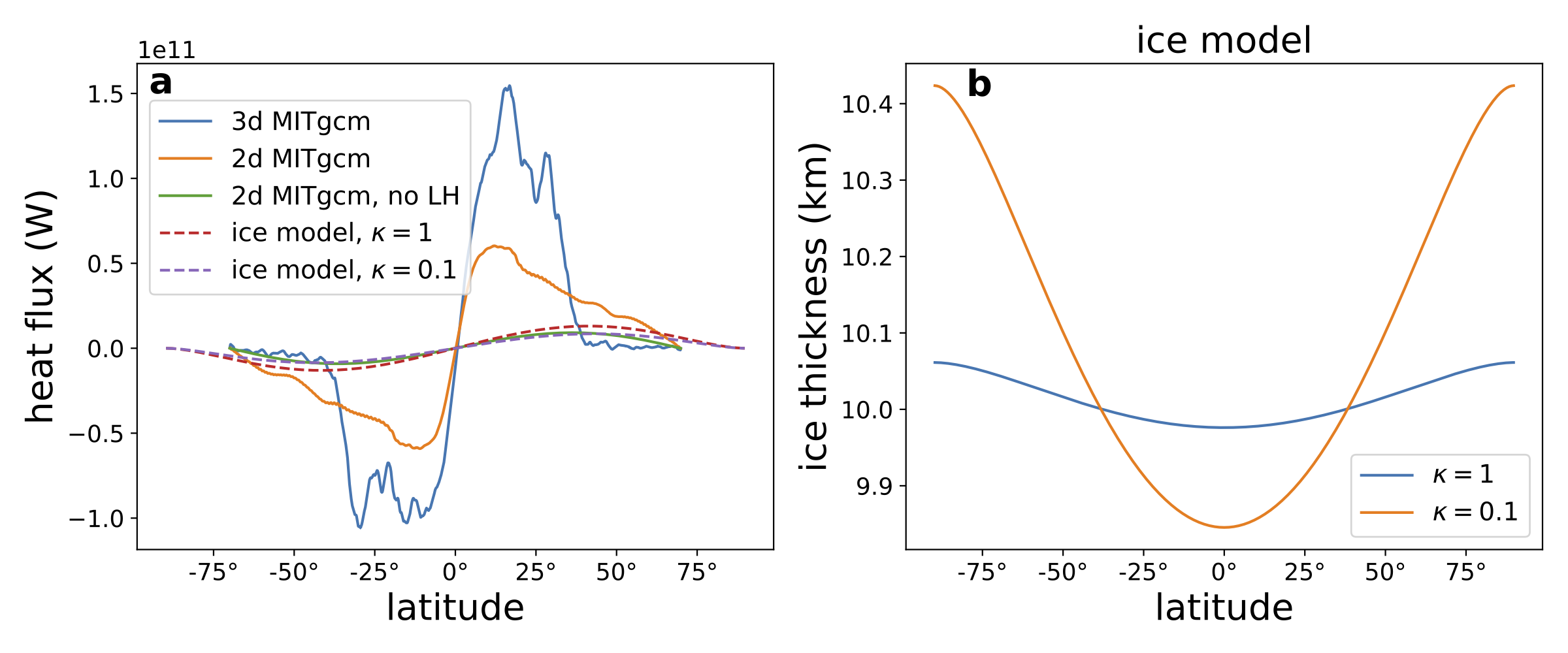}}
\caption{{\bf Meridional ocean heat transport and implications for ice thickness.} {\bf a} The oceanic meridional heat transport calculated using the 2d and 3d ocean simulations in this work (solid lines). The heat transport is positive northward, so that the heat transport is poleward in both hemispheres. The green solid line shows the meridional transport calculated from the difference between the geothermal heat flux and the diffusive heat flux through the ice (text). Also shown (thin dash lines) are estimates of ocean heat transport using the slab ocean model of ref. \cite{Ashkenazy-Sayag-Tziperman-2018:dynamics}, coupled to an ice flow model, for two oceanic eddy mixing values, $\kappa=1, 0.1$ m$^2$\,s$^{-1}$. {\bf b} Ice thickness (km) associated with the heat transport curves shown in {\bf a} (dashed lines) showing almost uniform ice thickness even for relatively small eddy mixing coefficient of $\kappa=1$ m$^2$\,s$^{-1}$.
  \label{fig:fig4}}
\end{figure}

\setcounter{figure}{0}

\clearpage\newpage

\section*{Methods}

\paragraph{Model description and configuration.} To investigate the dynamics of Europa's ocean, we used the state-of-the-art Massachusetts Institute of Technology ocean GCM (MITgcm\cite{Marshall-Adcroft-Hill-et-al-1997:finite,MITgcm-manual-github:mitgcm}). This model configuration used here employs the fully-nonlinear momentum equations for the ocean in height ($z$)-spherical coordinates, including the curvature terms, with a free surface. We use the non-hydrostatic option (rather than the more commonly used primitive equations which replace the vertical momentum equation with the hydrostatic approximation). The model simulates temperature and salinity using advection-diffusion equations, and uses a fully nonlinear equation of state relating them to the density field. The non-hydrostatic version of the MITgcm permits the use of the full Coriolis force, including the terms proportional to $2\Omega\cos\phi$ that are often neglected when the aspect ratio (depth over horizontal scale) is small. Europa's aspect ratio is relatively large, and this option is therefore very important for Europa's ocean, as was anticipated by\cite{Vance-Goodman-2009:oceanography}.

The MITgcm was also used to study diverse oceanic phenomena, as well as the dynamics of other planets and moons, including Jupiter\cite{Kaspi-2008:turbulent, Kaspi-Flierl-Showman-2009:deep}, Pluto\cite{Zalucha-Gulbis-2012:comparison, Zalucha-Michaels-2013:3d}, local convection on Europa\cite{Goodman-Lenferink-2012:numerical}, Triton\cite{Zalucha-Michaels-2013:3d} and hot Jupiter planets\cite{Showman-Fortney-Lian-et-al-2009:atmospheric, Parmentier-Fortney-Showman-et-al-2016:transitions}.  The model was used in the past to investigate the Snowball Earth events\cite{Ashkenazy-Gildor-Losch-et-al-2013:dynamics, Ashkenazy-Gildor-Losch-et-al-2014:ocean, Ashkenazy-Tziperman-2016:variability, Jansen-2016:turbulent}, which share many similarities with Europa's ice-covered ocean.

\paragraph{2d and 3d configurations.} We used two configurations, 2d and 3d, correspondingly. Both configurations extend almost globally in the meridional direction (70\degree{S} to 70\degree{N}). The depth of the ocean is set to 100 km. The lateral resolution of the 2d configuration is 1/12 of a degree (about 2.3 km in the meridional direction) and we use 50 vertical levels with a 2 km uniform resolution. The 3d simulation spans 30 degrees in the zonal direction with 1/24 of a degree resolution, using 100 vertical levels with varying vertical resolution, ranging from 25 m at the top of the ocean to 1164 m at the bottom. These horizontal and vertical 3d resolutions were found sufficient to resolve ocean eddy dynamics and convection processes. No-slip boundary conditions were assigned at the bottom and top (interface with the ice) of the ocean. The integration time steps were 7200 s and 400 s for the 2d and 3d configurations, respectively. Both simulations were ran for a sufficiently long time until statistical steady state was achieved.

\paragraph{Eddy coefficients, subgrid-scale representation.} The model uses \textit{explicit} eddy diffusion and viscosity coefficients that represent subgrid-scale mixing processes not resolved by the simulation. These are different from the eddy coefficients calculated below (Supplementary Fig.~12) which quantify the resolved eddy motions. The vertical explicit eddy viscosity and diffusion coefficients were set to $10^{-3}$ and $10^{-4}$ m$^2$ s$^{-1}$, respectively. The horizontal explicit eddy viscosity and tracer (thermal and haline) diffusion coefficients representing subgrid scale turbulent mixing are set to $50$ and $5$ m$^2$ s$^{-1}$, respectively, for the 2d configuration and $20$ and $2$ m$^2$ s$^{-1}$ respectively for the 3d configuration. The viscosity and diffusion coefficients are much larger than the molecular ones and are chosen to guarantee numerical stability. The eddy viscosity and diffusivity have different values, again a common practice in ocean modeling that is meant to allow using the smallest coefficients that lead to numerically stable results. The horizontal explicit eddy mixing coefficients are larger than the vertical ones, reflecting the different time scale of the subgrid scale turbulence represented by each.

These turbulent eddy coefficients can be formulated to be a function of the larger scale stratification and shear, as represented in present-day ocean model by the KPP parameterization, for example\cite{Large-Mcwilliams-Doney-1994:oceanic}. However, Europa's ocean is too different from Earth's to allow us to use such a parameterization, and we therefore set these coefficients to constant values. In addition, our non-hydrostatic simulations resolve the vertical convection plumes explicitly, even if marginally. We also do not use the Gent-McWilliams\cite{Gent-Mcwilliams-1990:isopycnal} eddy parameterization that is commonly used in Earth's ocean simulations, because the large-slope isopycnals that develop in the simulations violate the assumptions used to derive this parameterization.

\paragraph{Ice shell and bottom boundary condition.} Europa's icy shell is represented using the MITgcm shelf-ice package\cite{Losch-2008:modeling,MITgcm-manual-github:mitgcm} that enables the calculation of the ocean-ice freshwater and heat fluxes based on the surface ice temperature, ice thickness, and ocean temperature and salinity\cite{Losch-2008:modeling}. The forcing ice surface temperature was taken from\cite{Ashkenazy-2019:surface}. Ice flow and dynamical ice thickness are not included in the shelf-ice package, but, as demonstrated above based on\cite{Ashkenazy-Sayag-Tziperman-2018:dynamics}, the ice thickness can be assumed uniform due to the efficient meridional heat flux of the ocean (Fig.~\ref{fig:fig4}). A geothermal heating rate of 0.0496 W m$^{-2}$ is specified at the ocean bottom; the exact bottom heating is not known (estimated to be between 5-200 mW m$^{-2}$) and we use an intermediate estimate\cite{Hussmann-Spohn-Wieczerkowski-2002:thermal, Greenberg-Geissler-Hoppa-et-al-2002:tidal, Chen-Nimmo-Glatzmaier-2014:tidal} and performed sensitivity experiments to this value as shown in Supplementary Fig.~1-5. The internal heating leading to the geothermal heat flux is due to several sources: radiogenic heating of the metallic core and silicate mantle estimated at about 6--8 mW\,m$^{-2}$\cite{Barr-Showman-2009:heat}, and tidal heating of Europa's core\cite{Greenberg-Geissler-Hoppa-et-al-2002:tidal}. Tidal heating of the icy shell is not included explicitly, and tidal heating dissipation in the ocean is believed to be negligible\cite{Chen-Nimmo-Glatzmaier-2014:tidal}.

\paragraph{3-equation top boundary condition formulation.} We use the 3 equations-formulation of\cite{Losch-2008:modeling,MITgcm-manual-github:mitgcm} to calculate the freshwater and heat fluxes between the ice shell and the ocean. The formulation represents an unresolved boundary layer just under the ice where these exchanges occur. According to these equations, the heat balance of the boundary layer is,
\begin{equation}
  c_p\rho\gamma_T(T-T_f)+Lq=\rho_Ic_{p,I}\kappa\frac{T_f-T_s}{h},
  \label{eq:1}
\end{equation}
where the $c_p, c_{p,I}$ are the specific heat and water and ice, $\rho, \rho_I$ are the density of ice and ocean, $\gamma_T$ is the turbulent exchange coefficient of temperature, $T, T_s$ is the top of the ocean and surface of the ice temperatures, $\kappa$ is the diffusion constant through the ice, $h$ is the thickness of the ice, and $L$ the latent heat constant of fusion. The freezing temperature depends both on the pressure at the bottom of the ice (which is uniform in our case as the ice thickness is uniform) and on the boundary layer salinity, $S_b$, which is not uniform and calculated as part of the solution to the 3-equation model. As a result, even when the ice thickness and the freezing temperature are constant (or almost constant), the heat flux into the ice which appears in these equations as the above $\rho_Ic_{p,I}\kappa\frac{T_s-T_f}{h}$ term, is not uniform in latitude as the  temperature of the ice surface, $T_s$, strongly varies by tens of degrees with latitude.

The upper boundary conditions for the temperature and salinity-- that are behind the three-equation model used there, are essentially flux boundary conditions that take into account the effects of melting/ freezing on both the salt concentration and heat fluxes due to freezing/ melting and heat diffusion through the ice. The bottom boundary condition is again a flux boundary condition for both, specifying the geothermal heat flux and a no diffusive bottom flux for salt.

\paragraph{Differences in modeling strategy from previous studies of Europa's ocean.} Previous pioneering studies modeling Europa's ocean\cite{Soderlund-2019:ocean,Soderlund-Schmidt-Wicht-et-al-2014:ocean} were based on the MagIC model used for core magnetohydrodynamics applications\cite{Christensen-Wicht-2015:numerical}, and is therefore different from the currently used ocean model which has been developed to study ocean dynamics in particular. Among the differences: the MagIC model is pseudo spectral, while the MITgcm used here is finite volume, they use isotropic viscosity/diffusivity, while modern ocean studies use non-isotropic coefficients (different in the vertical and horizontal directions in spherical coordinates, representing different expected efficiency of mixing in the two directions). As mentioned in the paper body, the top and bottom boundary conditions are of prescribed temperature in the above studies, while we use a more self-consistent formulation involved a flux condition at the bottom and a 3-equation model at the bottom of the ice shelf. We explicitly represent the ice shelf, its melting, freezing and heat conduction -- all of which were ignored previously. These previous studies ignored salinity effects which are included here, and these salinity effects are found to dominate the density distribution as explained in the article. The previous studies used a linear equation of state relating the density to the temperature, while we use a full-complexity nonlinear equation of state relating density to temperature, salinity and pressure\cite{Jackett-Mcdougall-1995:minimal}. Finally, our resolution in both 2d and 3d is significantly higher than previously used.

\paragraph{Sensitivity to mean salinity, bottom heating and ice thickness.} Estimates of Europa's ocean salinity vary widely, from the ocean being nearly fresh to highly saline\cite{Zolotov-Kargel-2009:chemical}. Importantly, the magnetometer on the Europa Clipper may be able to estimate the mean ocean salinity\cite{Pappalardo-Prockter-Senske-et-al-2016:science, Howell-Pappalardo-2020:nasas}. The mean salinity affects the freezing temperature of ice and the ocean dynamics, as density variations are found in this work to be driven mostly by salinity gradients rather than temperature gradients. We have used a moderate salinity value of 50 gr/kg (ppt) as our default value. Given the uncertainty in this mean salinity value, we summarize in Supplementary Figs.~1-5 a set of 2d sensitivity tests that include mean salinity of 10$^{-6}$ ppt, 5 ppt, 10 ppt, 25 ppt. We also test the sensitivity of our results to the ice thickness, noting that the ice thickness should be in equilibrium with the geothermal bottom heat flux: as the thickness is changed, the diffusive heat flux through the ice changes, and a steady state requires the globally integrated heat flux through the ice to be equal to the integrated bottom heat flux. We used ice thickness values of 5 km (which is in equilibrium with an ocean bottom heat flux of 100 mW m$^{-2}$) and 15 km (corresponding to ocean bottom heat flux of 33 mW m$^{-2}$).

In all the simulations, we find that the coldest water is in the upper ocean, outside the tangent cylinder, as in our default simulation analyzed in the paper itself (Supplementary Fig.~1-5). The salinity is maximal outside the tangent cylinder (around the equator), except for the freshwater case (1st raw). The density is maximal outside the tangent cylinder for all simulation except the two lowest mean salinities (1st and 2nd rows) for which the anomaly of sea water leads to denser water at the bottom due to the bottom heating there. The top to bottom temperature difference is robustly at around 0.01\degree{C}, even when the bottom heat flux is changed from our default value. In all simulations the flow is westward except the bottom equatorial region for which superrotation is observed as discussed in the paper. The Taylor columns structure is clearly visible in the meridional velocity and is similar in all simulations, consistent with the results presented in the main text. Moreover, the structure of the temperature, salinity, and density fields is similar to structure of the those presented in the main text (Fig.~\ref{fig:fig-2d-results-tracers}). We conclude that the sensitivity simulations indicate the robustness of the results of the default experiment analyzed in the paper.

\paragraph{The dominant ocean momentum balance.} In order to explore the momentum balance of Europa's ocean, we use the output of the model that uses the full set of equations as explained above, but consider only those terms that are not negligible. We therefore consider the following equations, assuming zonal symmetry, and neglecting the small curvature terms, as well as vertical viscosity/diffusion terms which are very small due to the weakly stratified nature of Europa's ocean and due to the relatively small (in comparison to the horizontal one) vertical/diffusion viscosity coefficient. The momentum equations are then
\begin{linenomath*}
  \begin{align}
  u_t+\frac{1}{a} vu_\phi+wu_z-2\Omega \sin({\phi}) v+2\Omega \cos({\phi}) w
  & =\frac{\nu_h}{a^2\cos(\phi)} (\cos(\phi) u_\phi)_\phi
    +\nu_vu_{zz} \nonumber \\
  v_t+\frac{1}{a} vv_\phi+wv_z+2\Omega \sin(\phi) u
  &=-\frac{1}{a\rho_0}p_\phi+ \frac{\nu_h}{a^2\cos(\phi)} (\cos(\phi) v_\phi)_\phi
    +\nu_vv_{zz}\nonumber \\
  w_t+\frac{1}{a} vw_\phi+ww_z-2\Omega \cos(\phi) u &= -\frac{1}{\rho_0}{p_{\rm nh}}_z +\frac{\nu_h}{a^2\cos(\phi)}(\cos(\phi) w_\phi)_\phi
    +\nu_vw_{zz} \nonumber\\
  {p_{\rm hd}}_z&=-g\rho \nonumber \\
  p&=p_{\rm hd}+p_{\rm nh}
       \label{eq:full-MITgcm-equations-methods}
  \end{align}
\end{linenomath*}
where $\phi,z,t$ indicate the latitude, depth, and time, $u,v,w$ are the zonal, meridional and vertical velocities, $p_{\rm hd, nh}$ is the hydrostatic/nonhydrostatic pressure, $\rho$ is the density, $a$ is the radius of Europa, $\Omega$ is the rotation frequency, and $g$ is the gravity acceleration. The continuity equation assuming again zonal symmetry is,
\begin{linenomath*}
\begin{equation}
  \frac{1}{a \cos(\phi)} (v \cos(\phi) )_\phi+w_z=0.
\end{equation}
\end{linenomath*}
The different terms in the momentum and continuity equations are shown in Supplementary Figs.~14-16.
The Coriolis terms dominate the zonal horizontal momentum equation, followed by the horizontal viscosity term. In the meridional momentum equation, the balance is geostrophic: the horizontal Coriolis term balances the pressure gradient term, and the horizontal viscosity term is smaller yet not completely negligible.

While the model simulations shown in this work use a fully nonlinear equation of state\cite{Jackett-Mcdougall-1995:minimal} relating density to temperature, salinity and pressure, we note that a linearized equation can be written as $\rho=\rho_0(1-\alpha(T-T_0)+\beta(S-S_0))$ where $\alpha,\beta$ are the expansion coefficients mentioned in the main text. Because the temperature and salinity variations are very small, this linearized approximation is very accurate.

We now wish to explain the meridional symmetry of the velocity parallel to the axis of rotation, and the meridional anti-symmetry of the velocity perpendicular to the axis of rotation, as seen in Figs.~\ref{fig:fig-2d-results-tracers}b,c. The dominant terms in the zonal momentum equation are the Coriolis terms (Supplementary Fig.~15a,b) and the balance between them yields,
\begin{linenomath*}
\begin{equation}
  w\cos({\phi}){\approx}v\sin({\phi}).
  \label{eq:tan}
\end{equation}
\end{linenomath*}
This is more easily understood by writing the geostrophic approximation in cylindrical coordinates,  
\begin{linenomath*}
\begin{equation}
  2\Omega v_r\approx-\frac{1}{r\rho}\frac{\partial p}{\partial\theta},
\end{equation}
\end{linenomath*}
where now $v_r$ is the velocity perpendicular to the axis of rotation, written in terms of the spherical coordinate velocity field as $v_r=w\cos\phi-v\sin\phi$ and $\theta$ is the longitude. The 2d model configuration assumes zonal symmetry (no variations in $\theta$), so that the last equation implies $v_r\approx0$, exactly equivalent to (\ref{eq:tan}) in spherical coordinates.

The simple relation (\ref{eq:tan}), written as $w{\approx}v\tan(\phi)$, does not depend on any parameters and reflects the symmetry properties of $v,w$: The symmetry of $w$ is opposite of the symmetry of $v$ since $\tan({\phi})$ is anti-symmetric about the equator. Finally, the velocity parallel to the axis of rotation can be expressed in terms of the meridional and vertical velocities $v, w$ as $v_z=w\sin\phi+v\cos\phi$ and since $w\cos\phi\approx v\sin\phi$, $v_z=v/\cos\phi$, implying that $v_z$ is symmetric as $v$.

\paragraph{The spacing between the Taylor columns.} Based on Supplementary Figs.~14-16, 
the dominant terms in the zonal and meridional momentum equations near the top of the ocean, where we find viscosity to be non-negligible (\ref{eq:full-MITgcm-equations-methods}) are,
\begin{linenomath*}
  \begin{align}
-2\Omega \sin({\phi}) v  & =\frac{\nu_h}{a^2\cos(\phi)} (\cos(\phi) u_\phi)_\phi
\nonumber \\
2\Omega \sin(\phi) u
  &=-\frac{1}{a\rho_0}p_\phi+ \frac{\nu_h}{a^2\cos(\phi)} (\cos(\phi) v_\phi)_\phi.
       \label{eq:spacing1}
  \end{align}
\end{linenomath*}
The term $2\Omega \cos(\phi) w$ in the first equation is small in the upper 10 km or so of the ocean due to the no-normal flow conditions (Supplementary Fig.~15b). The viscosity term is generally smaller than the Coriolis term, especially in the interior (Supplementary Fig.~16a,b) in which geostrophy holds, $2\Omega \sin(\phi) \bar{u} =-\frac{1}{a\rho_0}p_\phi$, where $\bar{u}$ denotes the geostrophic zonal velocity. Subtracting the geostrophic balance from the fuller momentum equation (\ref{eq:spacing1}) and approximating the pressure with its geostrophic value throughout, we find,
\begin{linenomath*}
  \begin{equation}
    2\Omega \sin(\phi) (u-\bar{u})= \frac{\nu_h}{a^2\cos(\phi)} (\cos(\phi) v_\phi)_\phi.
    \label{eq:spacing2}
  \end{equation}
\end{linenomath*}
Assuming that we can neglect the meridional gradient of the geostrophic term is small, we can approximate $u_\phi\approx(u-\bar{u})_\phi$. This heuristic argument is supported by the smoother interior structure of the zonal velocity seen in Fig.~\ref{fig:fig-2d-results-vel}a. Eqs.~(\ref{eq:spacing1}) can now be written as,
\begin{linenomath*}
  \begin{align}
-2\Omega \sin({\phi}) v  & =\frac{\nu_h}{a^2\cos(\phi)} (\cos(\phi) \tilde{u}_\phi)_\phi
\nonumber \\
2\Omega \sin(\phi) \tilde{u}
  &= \frac{\nu_h}{a^2\cos(\phi)} (\cos(\phi) v_\phi)_\phi,
       \label{eq:spacing3}
  \end{align}
\end{linenomath*}
where $\tilde{u}=u-\bar{u}$. Using a complex variable $\alpha=\tilde{u}+iv$ the above equations can be written in terms of a single differential equation that holds near the top of the ocean where viscosity is non-negligible,
\begin{linenomath*}
  \begin{equation}
    i2\Omega \sin(\phi) \alpha= \frac{\nu_h}{a^2\cos(\phi)} (\cos(\phi) \alpha_\phi)_\phi\approx \frac{\nu_h}{a^2}\alpha_{\phi\phi},
    \label{eq:spacing4}
  \end{equation}
\end{linenomath*}
or
\begin{linenomath*}
  \begin{equation}
    \alpha_{\phi\phi}-ik_\phi^2\alpha=0; k_\phi^2=\frac{2\Omega a^2}{\nu_h}\sin(\phi).
    \label{eq:spacing5}
  \end{equation}
\end{linenomath*}
Given that the wave number $k_\phi^2$ is a slowly varying function of latitude (relative to the meridional scale of the Taylor columns), it can now be used to estimate the spacing between the Taylor columns. The corresponding wavelength in spherical coordinates is $\lambda=2\pi/k_\phi$. The distance of a given point, at a latitude $\phi$, along the ocean surface from the axis of rotation is given by $a\cos\phi$. The distance between two adjacent columns in the direction perpendicular to the axis of rotation, is therefore,
\begin{linenomath*}
  \begin{equation}
    d(\phi)=a[\cos(\phi-\lambda/2)-\cos(\phi+\lambda/2) ]=2a \sin(\phi)  \sin(\lambda/2)\approx a\lambda\sin(\phi).
    \label{eq:spacing6}
  \end{equation}
\end{linenomath*}
Or, more explicitly
\begin{linenomath*}
  \begin{equation}
    d(\phi)=\sqrt{2}\pi\sqrt{\frac{\nu_h}{\Omega}\sin(\phi)}.
    \label{eq:spacing7}
  \end{equation}
\end{linenomath*}
This approximation reproduces the functional behavior shown in Fig.~\ref{fig:taylor-spacing}a, although a factor that seems to be about $\pi$ is missing to be consistent with the numerical fit to the simulated distances. The above arguments are admittedly heuristic at best, yet suggest that eddy viscosity may indeed be at the heart of the process that sets the Taylor column spacing.

Following ref. \cite{Vance-Goodman-2009:oceanography}, we considered explaining the spacing between the Taylor columns through the Rhines scale, $L_\rho = \sqrt{2U/\beta}$. However, this scale does not seem to match the simulated spacing (more clearly shown by $u_{par}, u_{per}$, Figs.~\ref{fig:fig-2d-results-vel}b,c,e). At the equator $\beta=2.6\times 10^{-11}$ m$^{-1}$\,s$^{-1}$ and for a typical velocity of $U\sim 0.02$ m\,s$^{-1}$ the Rhines scale is $L_\rho \approx 40$~km, larger than the Taylor column spacing at the equator (between 10 and 20 km). For larger latitudes the Rhines scale becomes larger due to the division by $\beta$ which is proportional to the cosine of latitude. In contrast, the spacing we observe seems proportional to $\sqrt{\sin \phi}$ ($\phi$ is the latitude, Fig.~\ref{fig:taylor-spacing}), consistent with our revised explanation, and not to $\sqrt{1/\cos \phi}$ as predicted by the Rhines scale.

The scaling of ref. \cite{Zhang-Schubert-2000:magnetohydrodynamics} (their equation 23, based on their equation 22) in the case of rotating planet relates the Taylor column height ($D$) to its horizontal scale ($L$) as $L/D{\sim} \left(\frac{\nu_h}{2\Omega D^2}\right)^{1/3}$ so that $L{\sim} \left(\frac{D\nu_h}{2\Omega}\right)^{1/3}$. One can see that this cannot apply in our case for two reasons. First, there is a jump by a factor of two in the Taylor column height in our case across the tangent cylinder, but no jump is seen in the Taylor column spacing. Second, the Taylor column height in our Europa simulations first increases with latitude from the equator to the tangent cylinder and then decreases with latitude for higher latitudes. Their scaling would predict that the horizontal scale is therefore not monotonous in latitude (i.e., the spacing between the Taylor columns increases from the equator towards the tangent cylinder and then decreases towards the higher latitudes), in contrast to our numerical findings (Fig.~\ref{fig:taylor-spacing}) and to our own scaling arguments (given above) of monotonically increasing Taylor spacing from the equator towards the high latitudes.

\paragraph{Estimating eddy coefficients.} We estimate an horizontal eddy mixing coefficient, $\kappa_h$, that effectively represents the effects of ocean macro-turbulence, using time series of the zonal velocity at multiple grid points\cite{Hinze-1975:turbulence, Lemmin-1989:dynamics}. We use the auto-correlation function, $R(\tau)$, and the variance of the zonal velocity temporal anomaly as follows: $\kappa_h=\overline{u'^2_L}\int_0^\infty R(\tau)d\tau$, 
where $u_L$ is the Lagrangian zonal velocity, the overbar indicates mean over time while the prime indicates the temporal anomaly around this mean. When using the Eulerian velocity field, it is necessary to multiply $\kappa_h$ by a constant $\gamma$ which we choose to be $\gamma=4$\cite{Lemmin-1989:dynamics, Ashkenazy-Tziperman-2016:variability}. The estimated eddy parameterized viscosity coefficient is larger or equal to the estimated diffusion coefficient\cite{Lemmin-1989:dynamics, Ashkenazy-Tziperman-2016:variability}.

We find that the estimated diffusion coefficient depends on latitude, where for latitudes larger than 40\degree{} the diffusion coefficient is smaller than 40 m$^2$ s$^{-1}$ while of latitudes smaller than 40\degree{} it can reach a value larger than 1000 m$^2$ s$^{-1}$, depending on the latitude. The diffusion coefficient is also larger near the ocean bottom. We also estimated the diffusion/viscosity coefficient using an alternative approach based on the deviations of the zonal velocity from the zonal mean, and by estimating a characteristic length scale through the auto-correlation function in the zonal direction and multiplying it by the (zonal) standard deviation of the zonal velocity. We find diffusion/viscosity coefficients that are fairly similar to those found in the first method, of about 300 m$^2$ s$^{-1}$, again stronger at latitudes smaller than 40\degree and largest near the ocean bottom. Thus, following the above, a rough lower bound for the global mean eddy mixing coefficient is about 200 m$^2$ s$^{-1}$. Since the viscosity coefficient is usually larger than the diffusion coefficient, the above estimate for the viscosity coefficient of 300 m$^2$ s$^{-1}$ seems reasonable.

Surprisingly, these values are only an order of magnitude smaller than those estimated for Earth's ocean (e.g., $\sim$1000-5000 m$^2$ s$^{-1}$ in the tropical ocean\cite{Abernathey-Marshall-2013:global}) and of the same order as estimated for an Earth Snowball ocean\cite{Ashkenazy-Tziperman-2016:variability}, indicating that in spite of the lack of wind forcing and a direct solar forcing of the ocean, ocean eddies can develop from internal instabilities and are playing a dominant role in Europa's ocean dynamics and heat transport.

\paragraph{Scaling estimates of the role of rotation in convection dynamics.}
Previous studies\cite{Christensen-Aubert-2006:scaling} have suggested based on scaling arguments that rotation should affect the convection regime and therefore the top-to-bottom temperature difference. These scaling arguments assume a single component fluid (i.e., only temperature affecting the density), yet in our simulations the density variations are dominated by the salinity, while the temperature is close to the freezing temperature and therefore has only a small effect on the density. While the standard scaling is therefore not applicable in our case, we still calculate the modified and conventional nondimensional Rayleigh and Nusselt numbers based on the coefficients used in our Europa simulations and find that the conventional nondimensional numbers are much larger (by orders of magnitudes) than the modified numbers, indicating that rotation is not expected to play a role in the convection process. Specifically, the Prandtle number is, $Pr=\nu_v/\kappa_v$ ($\nu_v$ is the vertical viscosity coefficient and $\kappa_v$ is the vertical diffusion coefficient). The Ekman number is $E=\nu_v/({\Omega}D^2)$ ($\Omega$ is the rotation frequency and $D$ icy shell thickness). The thermal Ekman number is $Ek=\kappa_v/(\Omega{}D^2)=E/Pr$. The modified Rayleigh number in the presence of rotation is $Ra^*=(\alpha\Delta Tg)/(\Omega^2D)$, while the conventional Rayleigh number is $Ra=Ra^*/(Ek\,E)$. Similarly, the conventional Nusselt number is $Nu=qD/(\rho c_p\kappa_v\Delta T$) where $q$ is the heat flux, while the modified Nusselt number in the presence of rotation is $Nu^*=(Nu-1)Ek$. Calculating these nondimensional numbers for values corresponding to our Europa simulations, we find, $\Omega=2\times10^{-5}$ s$^{-1}$, $\kappa_v=1\times10^{-4}$ m$^2$s$^{-1}$, $\nu_v=1\times10^{-3}$ m$^2$s$^{-1}$, $D=10^{5}$, $Pr=10$, $g=1.314$ m\~s$^{-2}$, $q=0.05$ Wm$^{-2}$, $\alpha\Delta T=\Delta\rho/\rho=2\times10^{-6}$, $E=5\times10^{-9}$, $Ek=5\times10^{-10}$, $Ra^*=(2\times10^{-6})/(4\times10^{-10}\times10^{5})=0.05$, $Ra=0.05/(5\times10^{-10}{\times}5\times10^{-9})=2\times10^{16}$, $Nu=0.05{\times}10^{5}/(10^{3}{\times}4\times10^{3}{\times}10^{-4}\times0.01)=1250$, $Nu^*=(Nu-1)Ek=6\times10^{-7}$. Thus, the conventional Rayleigh and Nusselt numbers are much larger than the modified ones, indicating that the rotation is not expected to play a major role in the convection process. It is possible to define a Nusselt number that depends on density variations and not temperature. It is $Nu=q/(g\kappa_v\Delta\rho)$, which yields $Nu=0.05/(10^{-4}{\times}2\times10^{-4})=2.5\times10^{6}$, so that again the rotation is not expected to be a major factor, even when taking salinity changes into account.

\paragraph{Data availability.} The datasets generated during and/or analysed during the current study are available from the corresponding author on reasonable request. The model's setup files are available in the OSF repository, \href{http://OSF.IO/SVXBQ}{http://OSF.IO/SVXBQ}, DOI \href{https://doi.org/10.17605/OSF.IO/SVXBQ}{https://doi.org/10.17605/OSF.IO/SVXBQ}.

\paragraph{Code availability.} The reported results were generated using the MITgcm code which can be downloaded from \href{https://github.com/MITgcm/MITgcm}{https://github.com/MITgcm/MITgcm} or \href{https://doi.org/10.5281/zenodo.1409237}{https://doi.org/10.5281/zenodo.1409237}.

\paragraph{ACKNOWLEDGMENTS.}
We thank Francis Nimmo and Yohai Kaspi. E.T. thanks the Weizmann Institute for its hospitality during parts of this work. Y.A.~was funded by U.S-Israel Binational Science Foundation (BSF grant number 2018152) for financial support. E.T.~was funded by the National Aeronautics and Space Administration Habitable Worlds programme (grant FP062796-A/NNX16AR85G).

\bigskip
\paragraph{Author contribution.}
YA and ET took part in all stages of the work.

\bigskip
\paragraph{Competing Interests.} The authors declare no competing interests.


\newpage

\renewcommand{\figurename}{Supplementary Figure}
\setcounter{figure}{0}

\begin{figure}
  \centerline{\includegraphics[width=0.85\linewidth]{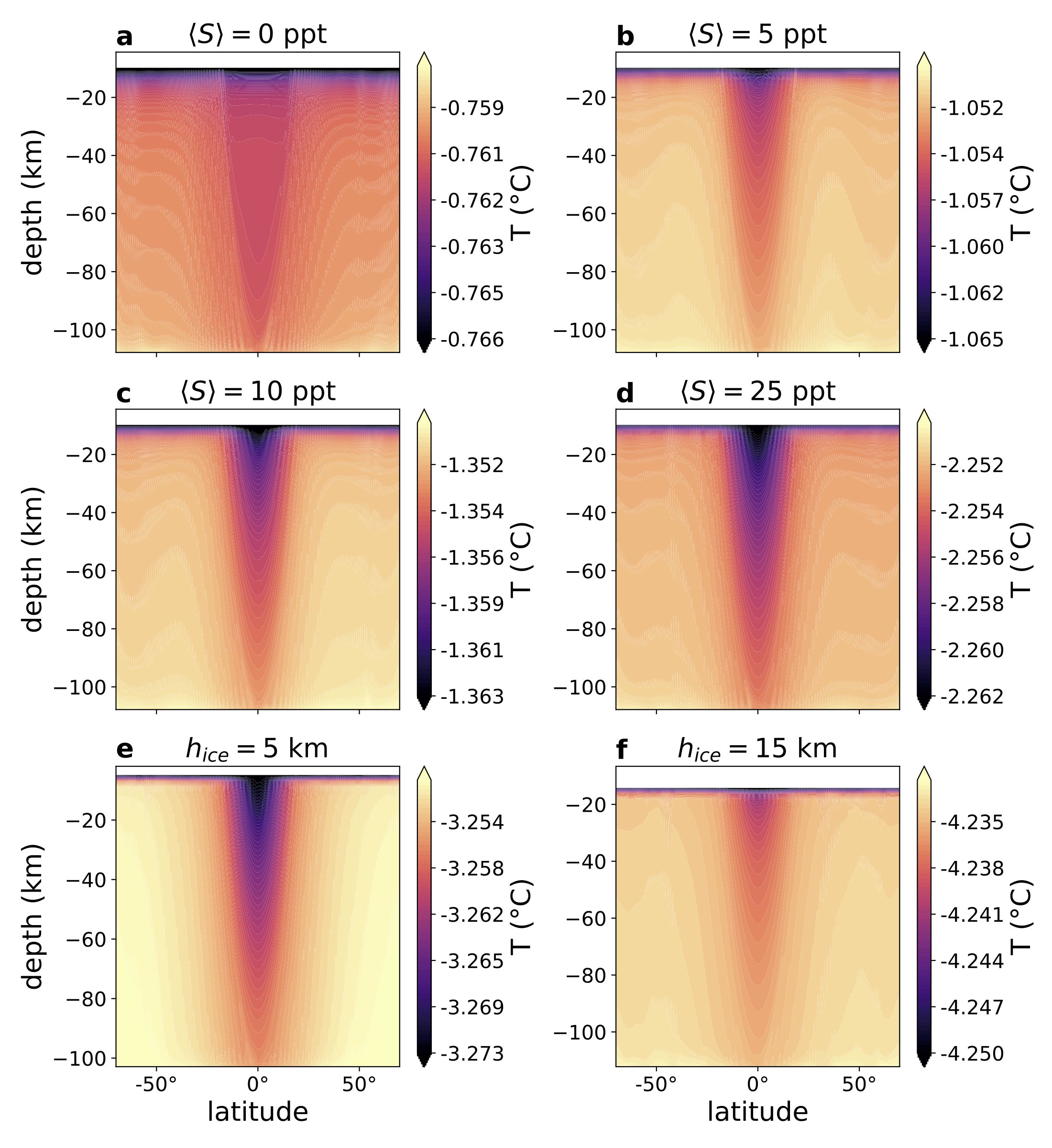}}
\vspace*{-92ex}  
\begin{center}
\Large{Supplementary Figures}
\end{center}
\vspace*{77ex}
\caption{ {\bf Sensitivity tests--temperature.} 2d  latitude-depth snapshots of temperature, $T$, for mean salinity of {\bf a} 10$^{-6}$ ppt, {\bf b} 5 ppt, {\bf c} 10 ppt, {\bf d} 25 ppt, and ice thickness of {\bf e} 5 km (corresponding to ocean bottom heat flux of 100 mW m$^2$), and {\bf f} 15 km (corresponding to ocean bottom heat flux of 33 mW m$^2$). Note that in panel a, with the vanishing mean salinity, the source of the bottom dense water is the bottom heating combined with the water anomaly at this range of temperatures, which leads to a density increase with heating.
  \label{fig:2d-sensitivity-T}}
\end{figure}

\begin{figure}
\centerline{\includegraphics[width=\linewidth]{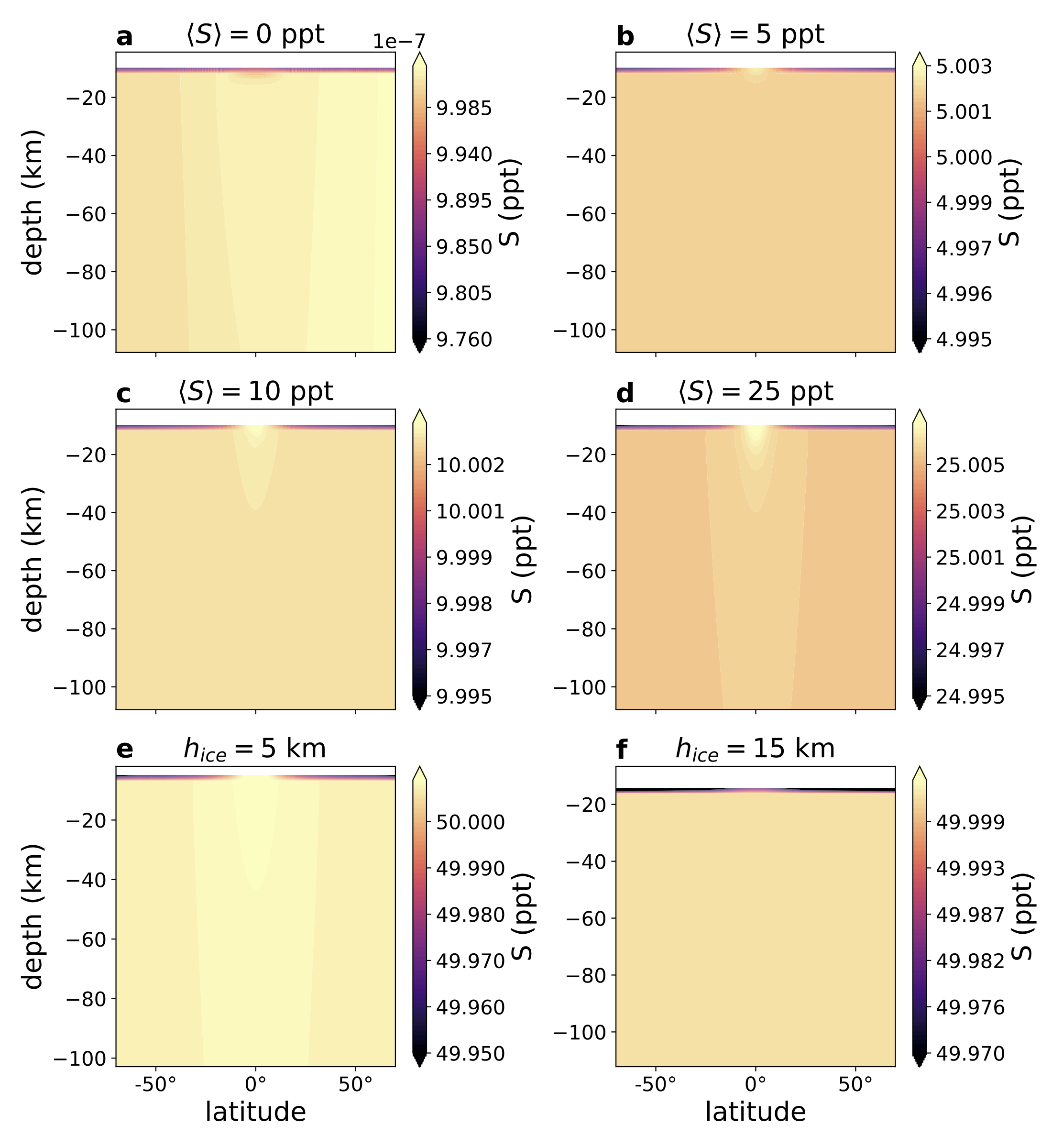}}
\caption{ {\bf Sensitivity tests--salinity.} Same as Supplementary Fig. \ref{fig:2d-sensitivity-T} for salinity, $S$.
  \label{fig:2d-sensitivity-S}}
\end{figure}

\begin{figure}
\centerline{\includegraphics[width=\linewidth]{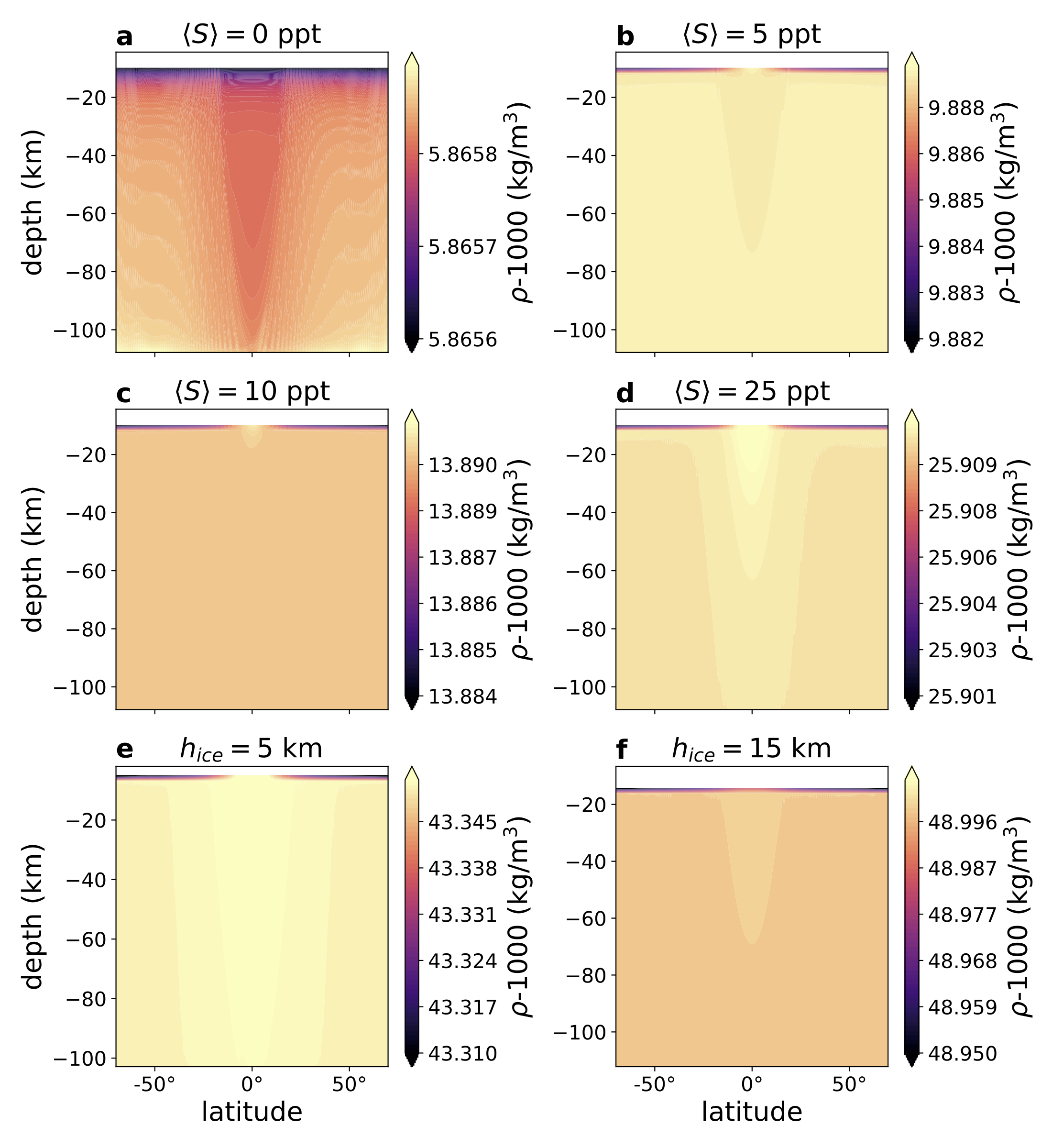}}
\caption{ {\bf Sensitivity tests--density.} Same as Supplementary Fig. \ref{fig:2d-sensitivity-T} for density, $\rho$.
  \label{fig:2d-sensitivity-rho}}
\end{figure}

\begin{figure}
\centerline{\includegraphics[width=\linewidth]{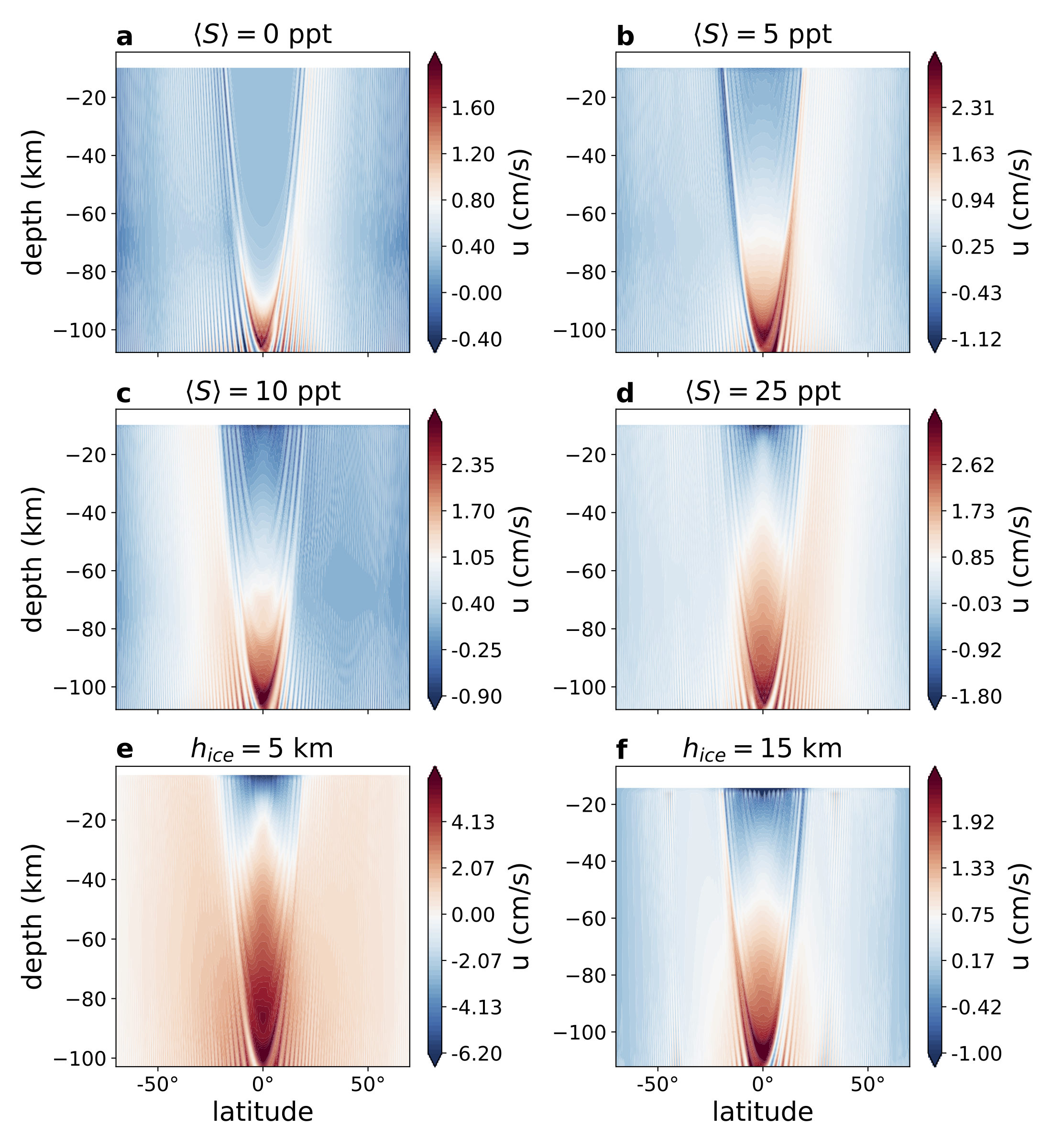}}
\caption{ {\bf Sensitivity tests--zonal velocity.} Same as Supplementary Fig. \ref{fig:2d-sensitivity-T} for zonal velocity, $u$.
  \label{fig:2d-sensitivity-U}}
\end{figure}

\begin{figure}
\centerline{\includegraphics[width=\linewidth]{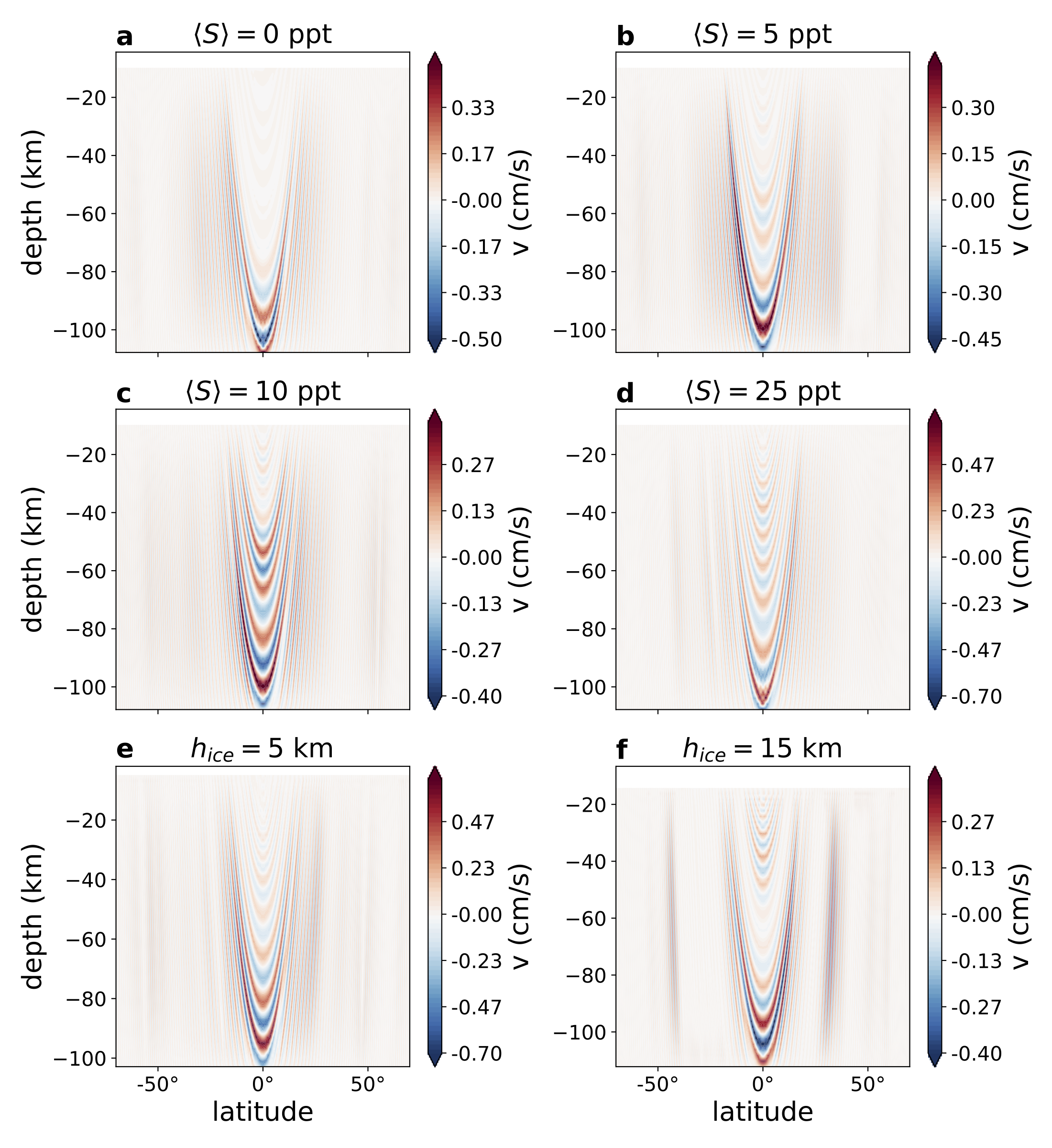}}
\caption{ {\bf Sensitivity tests--meridional velocity.} Same as Supplementary Fig. \ref{fig:2d-sensitivity-T} for meridional velocity, $v$.
  \label{fig:2d-sensitivity-V}}
\end{figure}

\begin{figure}
\centerline{\includegraphics[width=0.75\linewidth]{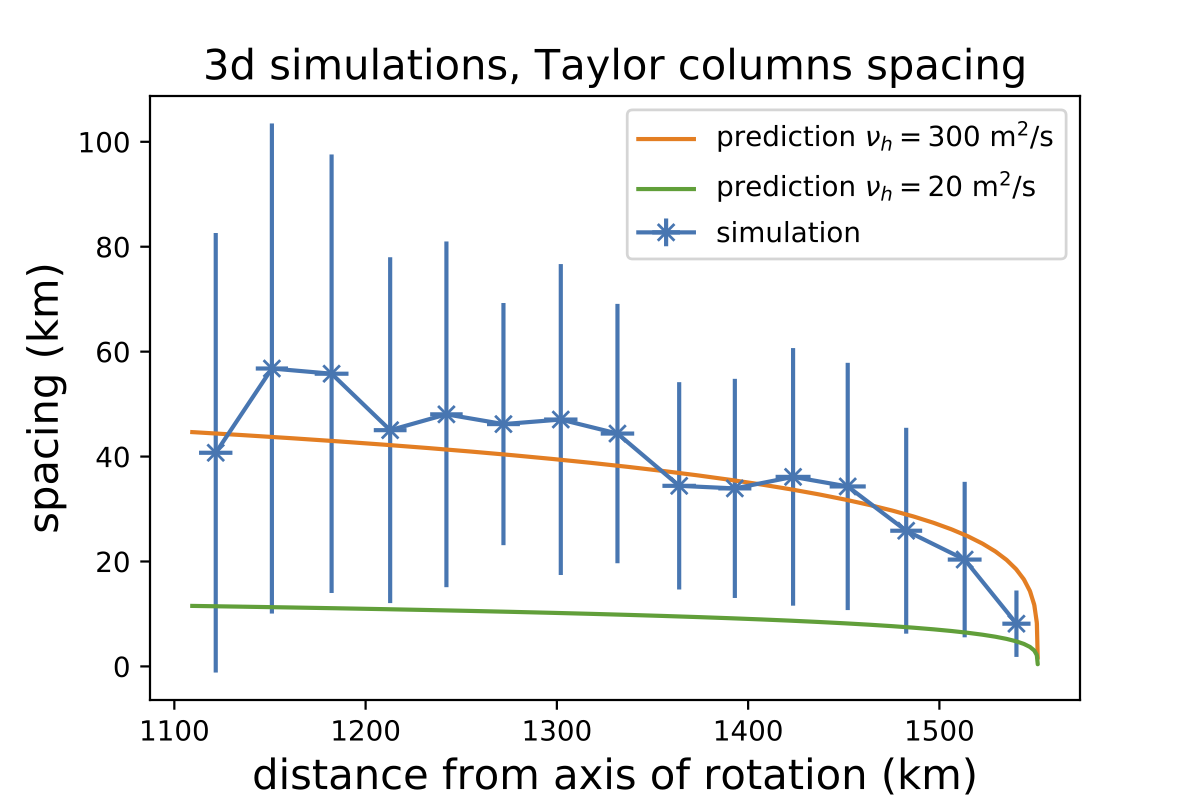}}
\caption{
  {\bf Taylor columns spacing based on the 3d simulation.} The distance between the Taylor columns as a function of the distance from the axis of rotation (in km, blue asterisks). The results are shown for a snapshot at a single time, and the spacing between the columns is calculated for each longitude grid point where then the data was binned using 60 km interval; the std  is shown by the vertical bars. The orange line represents the predicted spacing with an eddy viscosity coefficient of $\nu_h=300$ m$^2$s$^{-1}$ which fits the numerical values. This eddy coefficient is 15 times larger than the explicit viscosity coefficient used in the numerical simulation (green line), suggesting that the eddy viscosity coefficient due to the explicitly resolved eddy motions in the 3d simulation is 15 times larger than the explicit one.
\label{fig:figS5}}
\end{figure}

\begin{figure}
\centerline{\includegraphics[width=\linewidth]{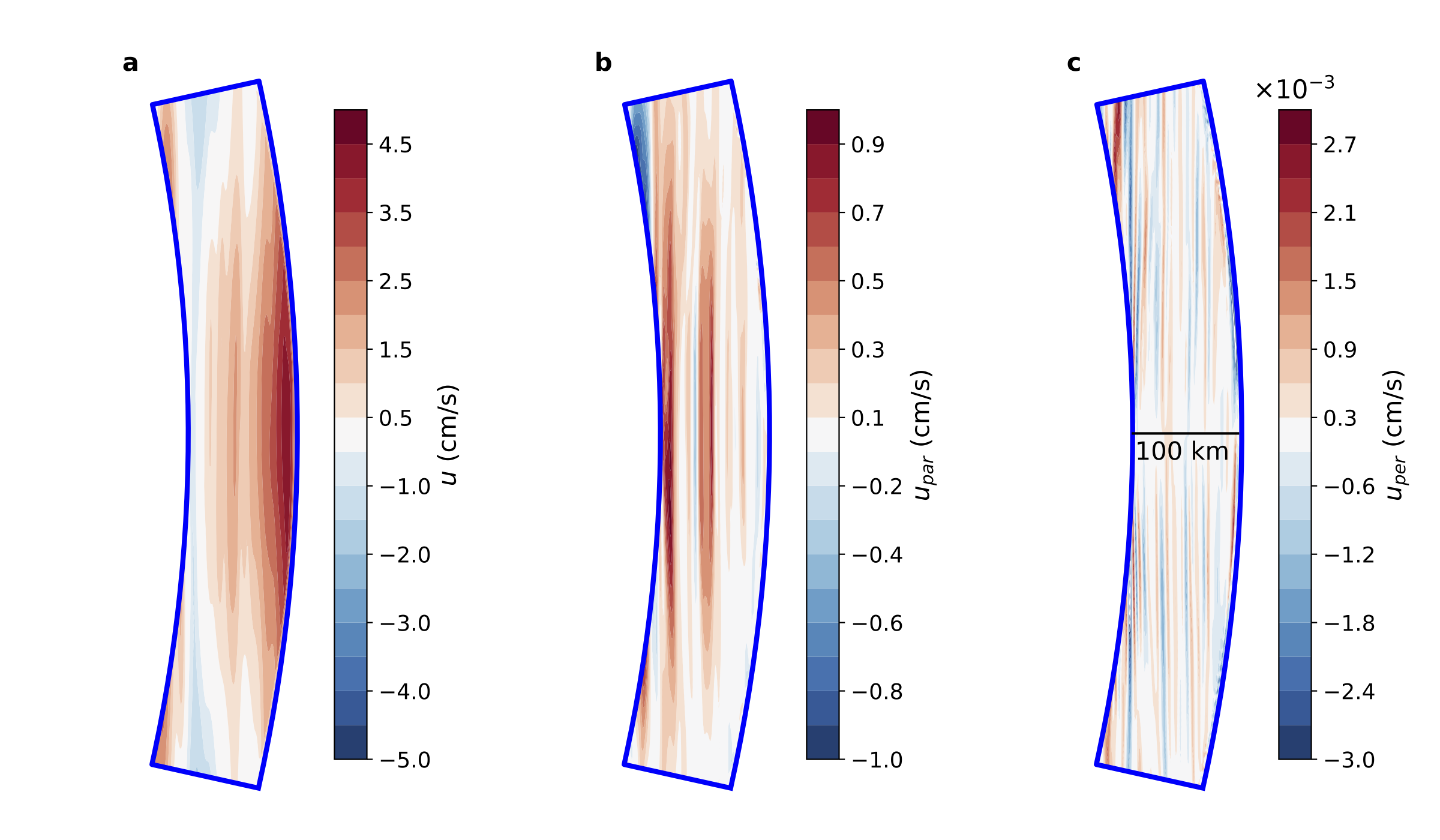}}
\caption{{\bf Taylor columns in the 3d simulation.} The velocity components are shown in a spherical projection to supplement the depth-latitude projection in Fig.~5 
  in the main text: Zonal mean of {\bf a} zonal velocity, $u$, {\bf b} velocity parallel to the axis of rotation, $u_{par}$, and {\bf c} velocity perpendicular to the axis of rotation, $u_{per}$. The latitudinal extent of the plots is from 12.5\degree{S} to 12.5\degree{N} and the depth extent is 100 km.  
\label{fig:Fig-3d-Taylor-columns-sphere}}
\end{figure}

\begin{figure}
\centerline{\includegraphics[width=\linewidth]{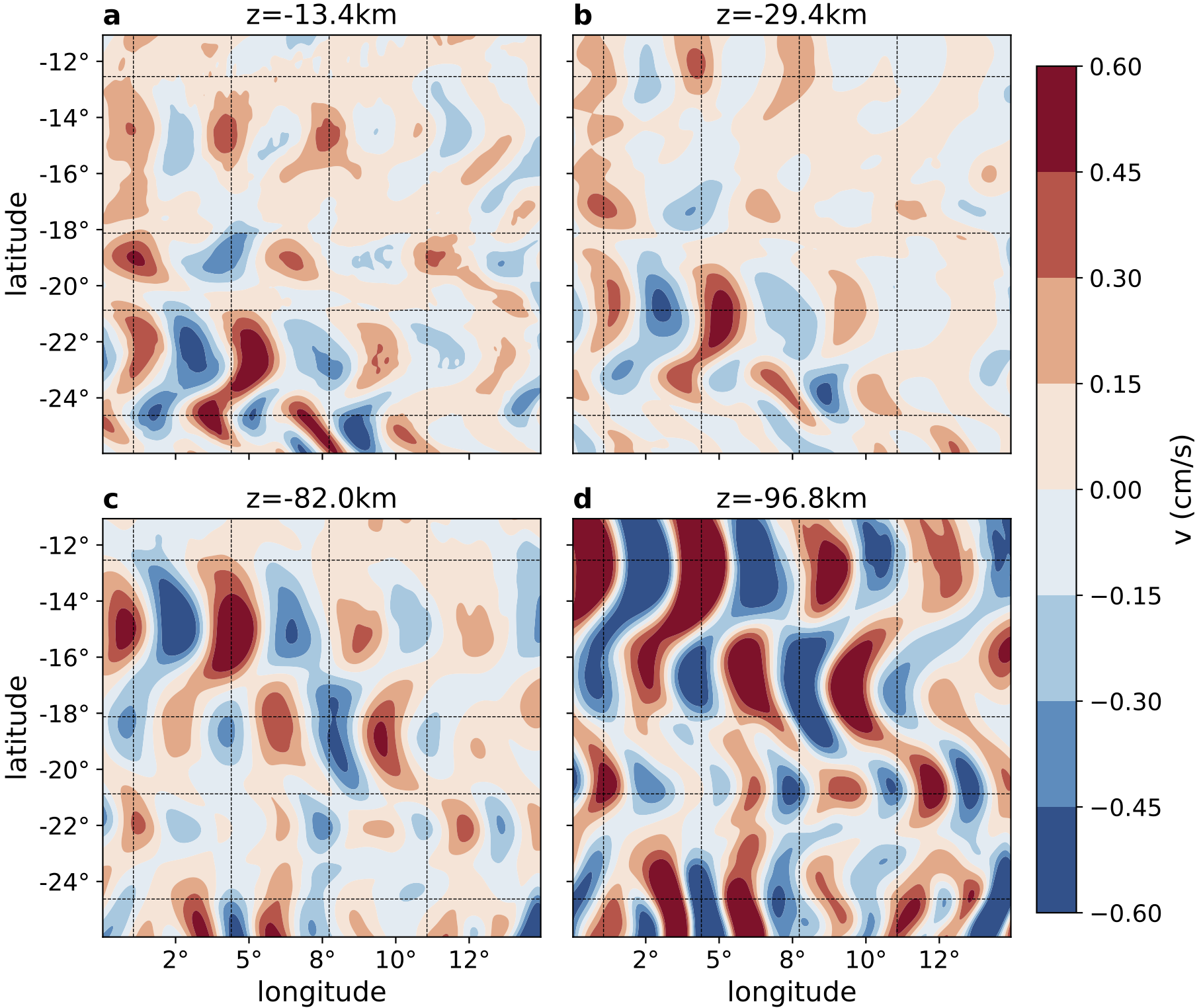}}
\caption{{\bf Structure of 3d Taylor columns--longitude-latitude sections.} Longitude-latitude plots of the meridional velocity, $v$, (in cm s$^{-1}$) at different depth of {\bf a} $z=-13.4$ km, {\bf b} $z=-29.4$, {\bf c} $z=-82$ km, and {\bf d} $z=-96.8$ km. The vertical dashed lines indicate the zonal sections plotted in Supplementary Fig. \ref{fig:fig-3d-v-levels-e-h} while the horizontal dashed lines indicate the zonal sections plotted in Supplementary Fig. \ref{fig:fig-3d-v-levels-i-l}. The dotted curved line indicates a line parallel to the axis of rotation. 
\label{fig:fig-3d-v-levels-a-d}}
\end{figure}

\begin{figure}
\centerline{\includegraphics[width=\linewidth]{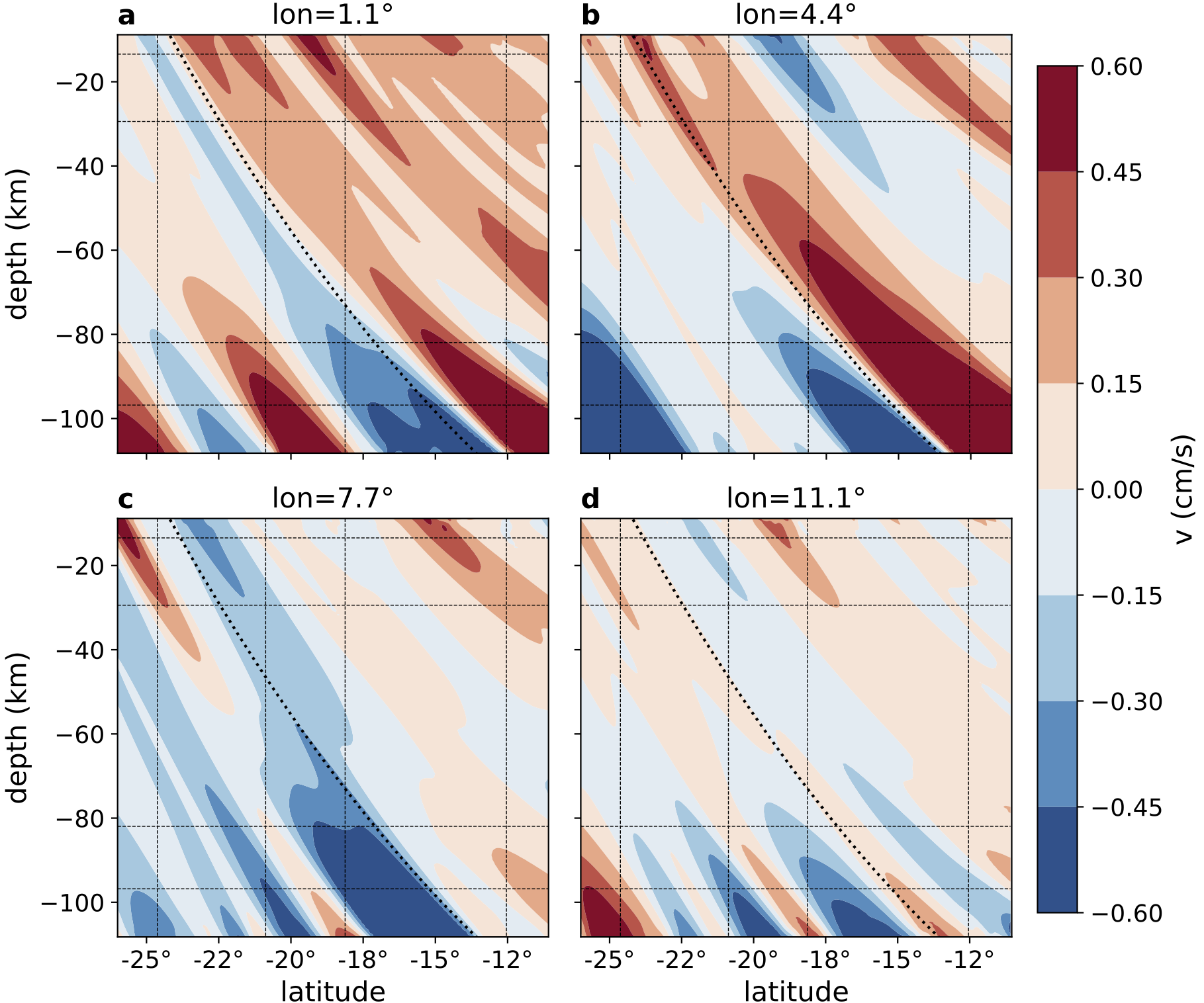}}
\caption{{\bf Structure of 3d Taylor columns--latitude-depth sections.} Longitude-depths plots of the meridional velocity, $v$, (in cm s$^{-1}$) at different longitudes of {\bf a} 1.1\degree{}, {\bf b} 4.4\degree{}, {\bf c} 7.7\degree{}, and {\bf d} 11.1\degree{}. The vertical dashed lines indicate the meridional sections plotted in Supplementary Fig. \ref{fig:fig-3d-v-levels-i-l} while the horizontal dashed lines indicate the depth sections plotted in Supplementary Fig. \ref{fig:fig-3d-v-levels-a-d}. The dotted curved line indicates a line parallel to the axis of rotation. 
\label{fig:fig-3d-v-levels-e-h}}
\end{figure}

\begin{figure}
\centerline{\includegraphics[width=\linewidth]{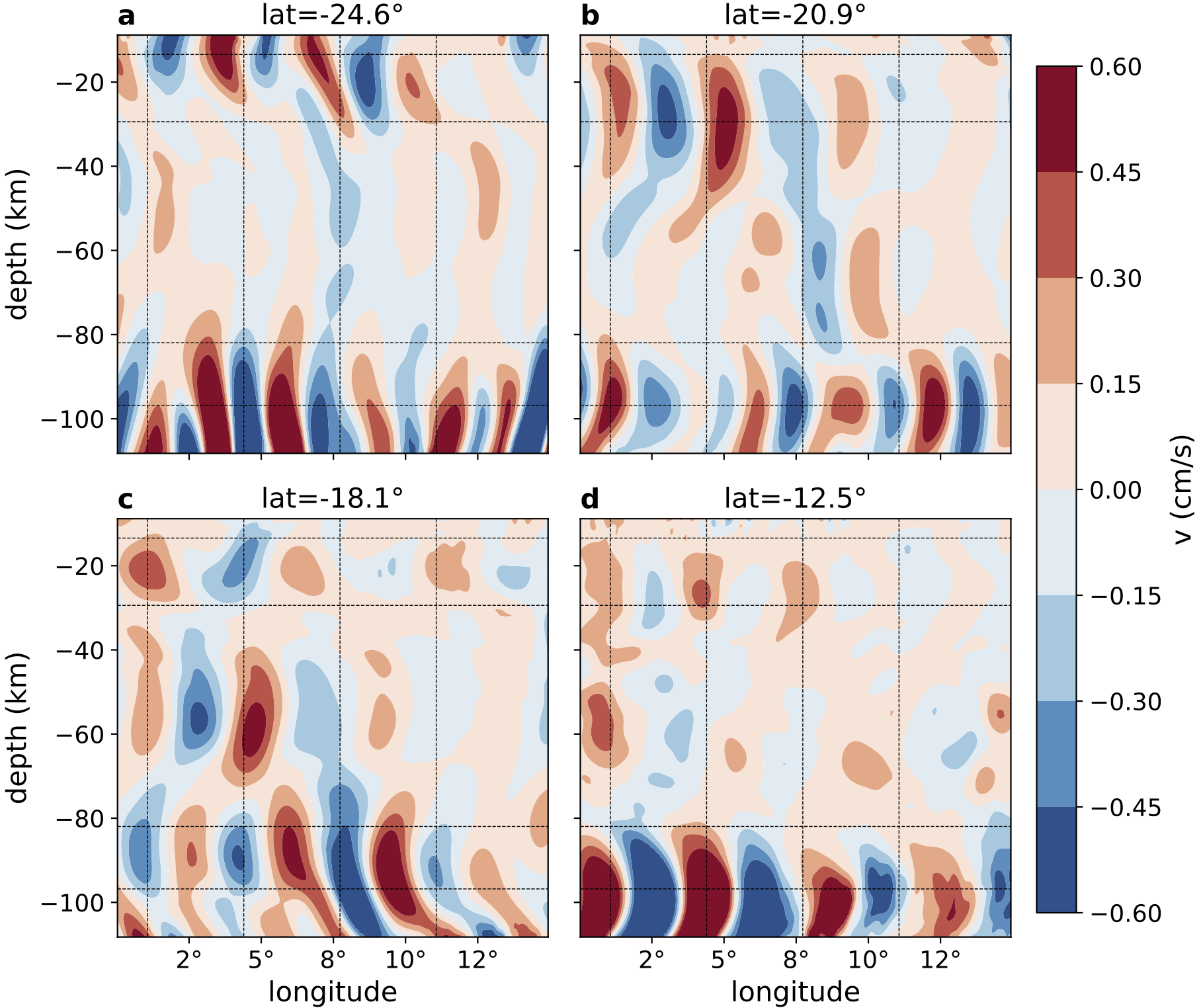}}
\caption{{\bf Structure of 3d Taylor columns--longitude-depth sections.} Longitude-depth plots of the meridional velocity, $v$, (in cm s$^{-1}$) at different latitudes of {\bf a} -24.6\degree{}, {\bf b} -20.9\degree{}, {\bf c} -18.1\degree{}, and {\bf d} -12.5\degree{}. The vertical dashed lines indicate the zonal sections plotted in Supplementary Fig. \ref{fig:fig-3d-v-levels-e-h} while the horizontal dashed lines indicate the depth sections plotted in Supplementary Fig. \ref{fig:fig-3d-v-levels-a-d}. The dotted curved line indicates a line parallel to the axis of rotation. 
\label{fig:fig-3d-v-levels-i-l}}
\end{figure}

\begin{figure}
\centerline{\includegraphics[width=\linewidth]{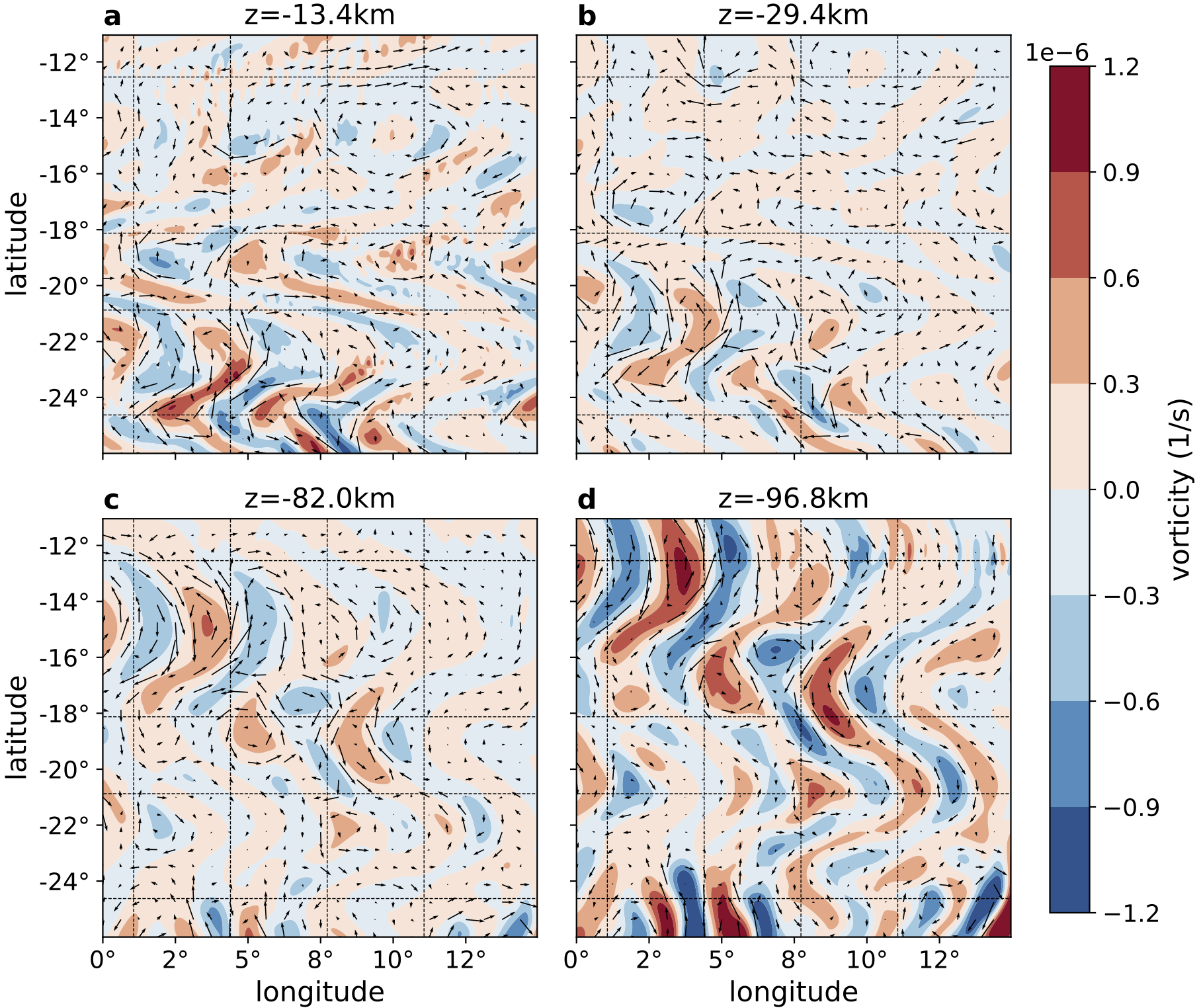}}
\caption{{\bf Structure of 3d Taylor columns--longitude-latitude section of vorticity.} Same as Supplementary Fig. \ref{fig:fig-3d-v-levels-a-d} for vorticity (in s$^{-1}$).
\label{fig:fig-3d-v-levels-m-p}}
\end{figure}

\begin{figure}
\centerline{\includegraphics[width=0.7\linewidth]{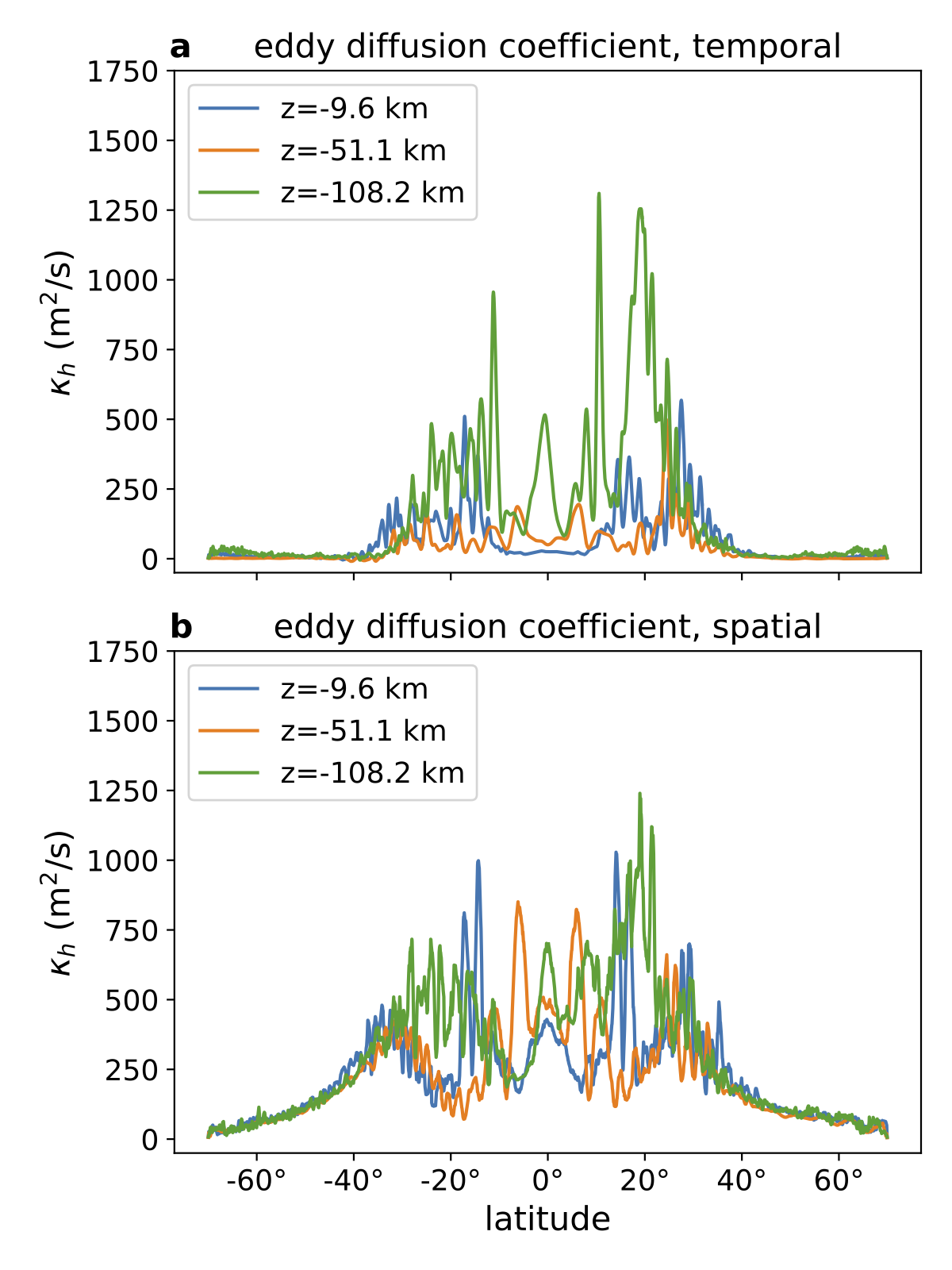}}
\caption{ {\bf Estimating the eddy coefficients.} The estimate is based on the 3d simulation using the {\bf a} temporal auto-correlation function and {\bf b} spatial auto-correlation function. The zonal mean at the top ($z=-9.6$ km, blue), middle ($z=-51.1$ km, orange), and bottom ($z=-108.2$ km, green) of the ocean is plotted versus latitude where the estimated diffusion coefficient, $\kappa_h$, is significantly smaller at the high latitudes. The estimated eddy viscosity coefficient is equal to or larger than the eddy diffusion coefficient.
\label{fig:figS3}}
\end{figure}

\begin{figure}
\centerline{\includegraphics[width=0.9\linewidth]{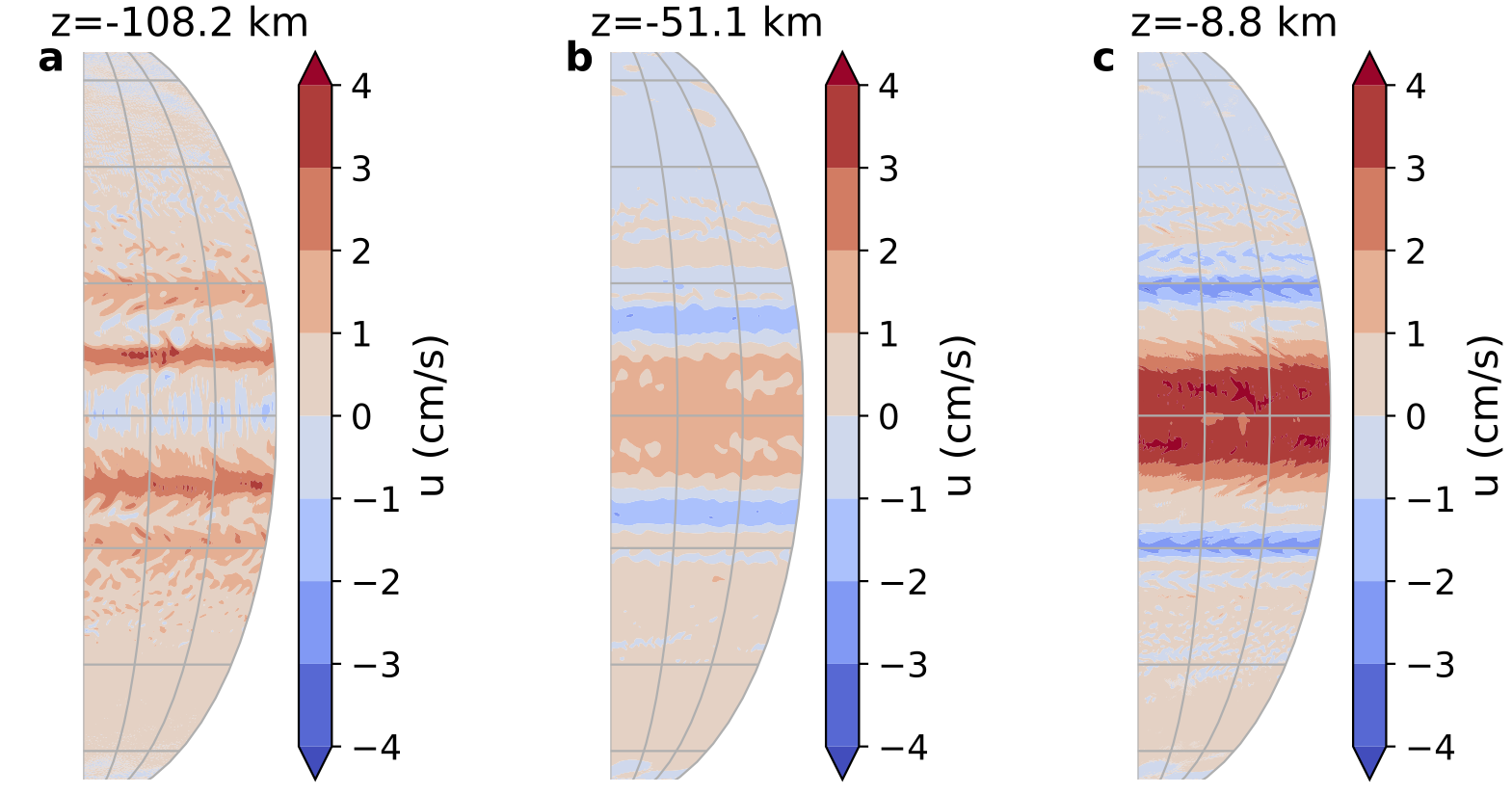}}
\caption{ {\bf Zonal velocity, $u$, in 3d simulation.} Shown in a spherical (longitude-latitude) projection at the {\bf a} bottom ($z=-108.2$ km), {\bf b} middle ($z=-51.1$ km), and {\bf c} top ($z=-8.8$ km) of the ocean. The grid line spacing is 10\degree{} in the zonal direction and 20\degree{} in the meridional direction. The figure depicts a ``Jupiter-like'' structure of alternating zonal jets as was previously predicted\cite{Vance-Goodman-2009:oceanography}.
\label{fig:Fig-3d-u}}
\end{figure}

\begin{figure}
\centerline{\includegraphics[width=\linewidth]{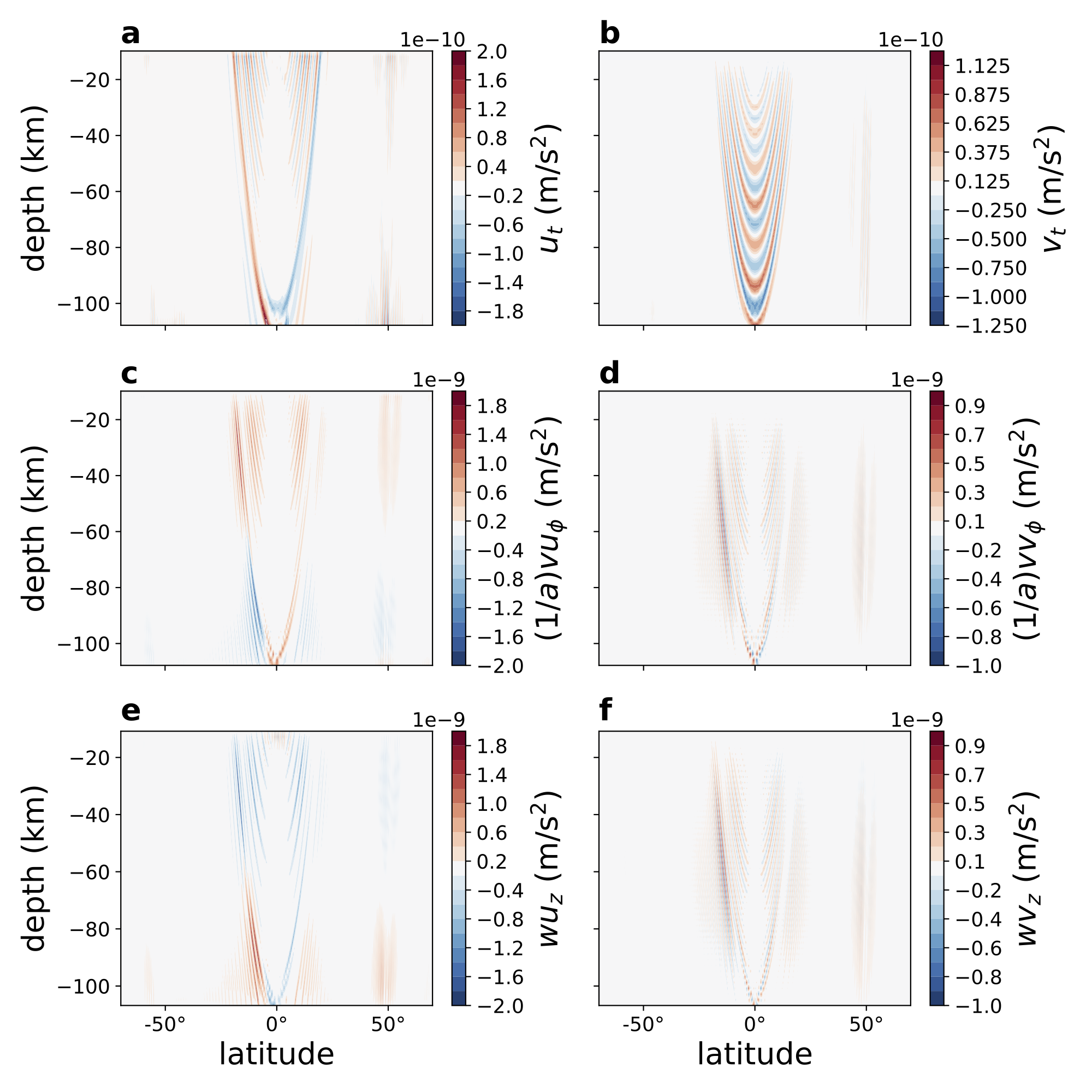}}
\caption{
  {\bf Time and advection terms in the momentum equations.} Snapshots of the: {\bf a} time derivative of the zonal velocity, $u_t$, {\bf b} time derivative of the zonal velocity, $v_t$, {\bf c} meridional advection of the zonal velocity, $\frac{1}{a}vu_\phi$, {\bf d} meridional advection of the meridional velocity, $\frac{1}{a}vv_\phi$, {\bf e} vertical advection of the zonal velocity, $\frac{1}{a}wu_z$, and {\bf f} vertical advection of the moridional velocity, $\frac{1}{a}wv_z$. $\phi, z, t$ are the meridional, vertical, and time coordinates, $u,v,w$ are the zonal, meridional, and meridional velocities, and $a$ is the radius of Europa. 
\label{fig:fig-2d-terms-dt-advection}}
\end{figure}

\begin{figure}
\centerline{\includegraphics[width=0.9\linewidth]{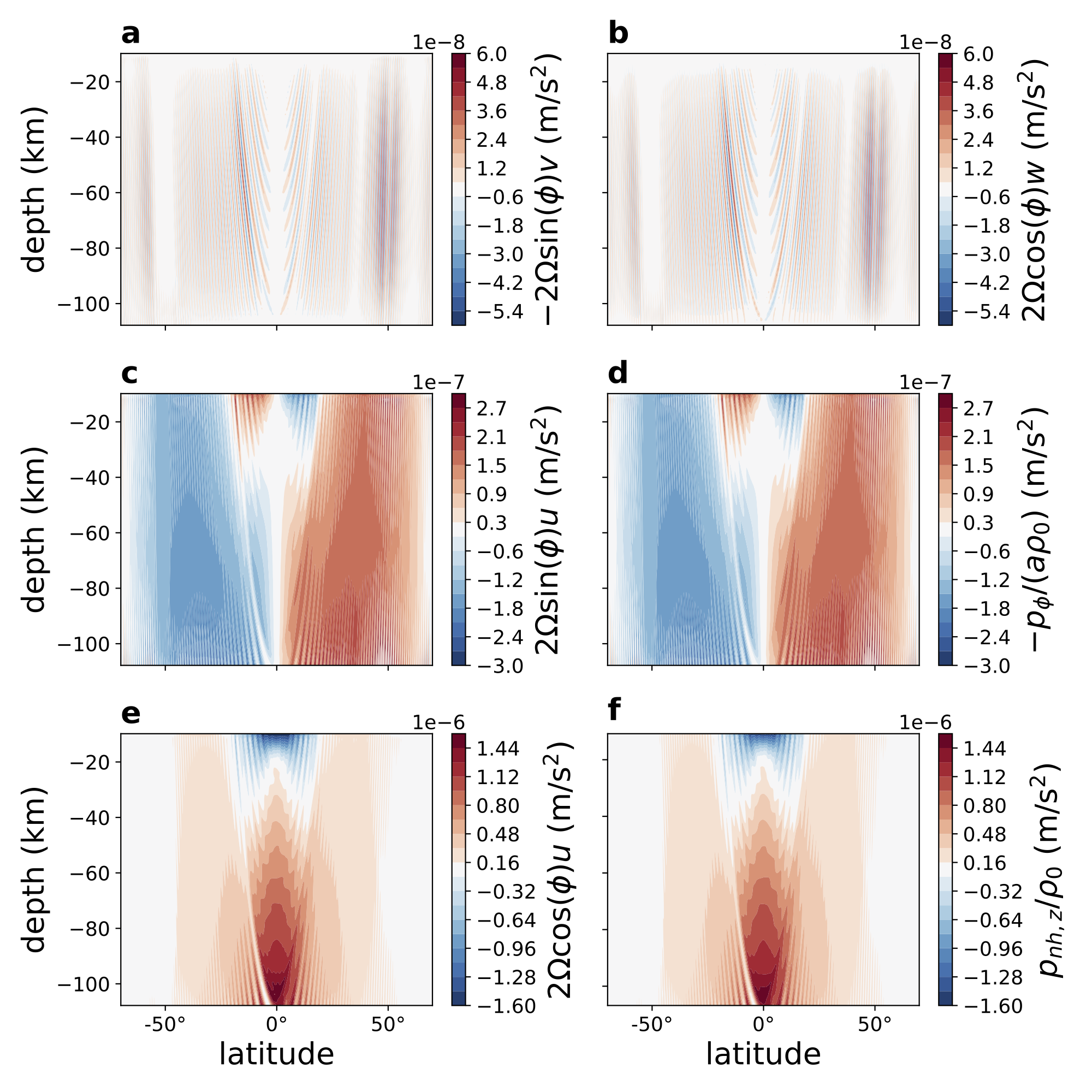}}
\caption{
  {\bf Coriolis and pressure terms in the momentum equations.} Snapshots of the: {\bf a} Coriolis term, $-2\Omega\sin(\phi)v$, {\bf b} co-Coriolis term, $2\Omega\cos(\phi)w$, {\bf c} Coriolis term, $2\Omega\sin(\phi)u$, {\bf d} meridional pressure gradient, $-\frac{1}{a\rho_0}p_\phi$, {\bf e} co-Coriolis term, $2\Omega\cos(\phi)u$, and {\bf f} vertical (non-hydrostatic) pressure gradient term, $\frac{1}{\rho_0}p_{nh,z}$.  Panels {\bf e} and {\bf f} depict the most dominant terms in the vertical momentum equation. $\phi, z$ are the meridional and vertical coordinates, $u,v,w$ are the zonal, meridional, and meridional velocities, $p, p_{nh}$ are the total and non-hydrostatic pressures, $a$ is the radius of Europa, $\Omega$ is the rotation rate of Europa, and $\rho_0$ is the reference density of Europa's ocean. 
\label{fig:fig-2d-terms-coriolis-p}}
\end{figure}

\begin{figure}
\centerline{\includegraphics[width=0.65\linewidth]{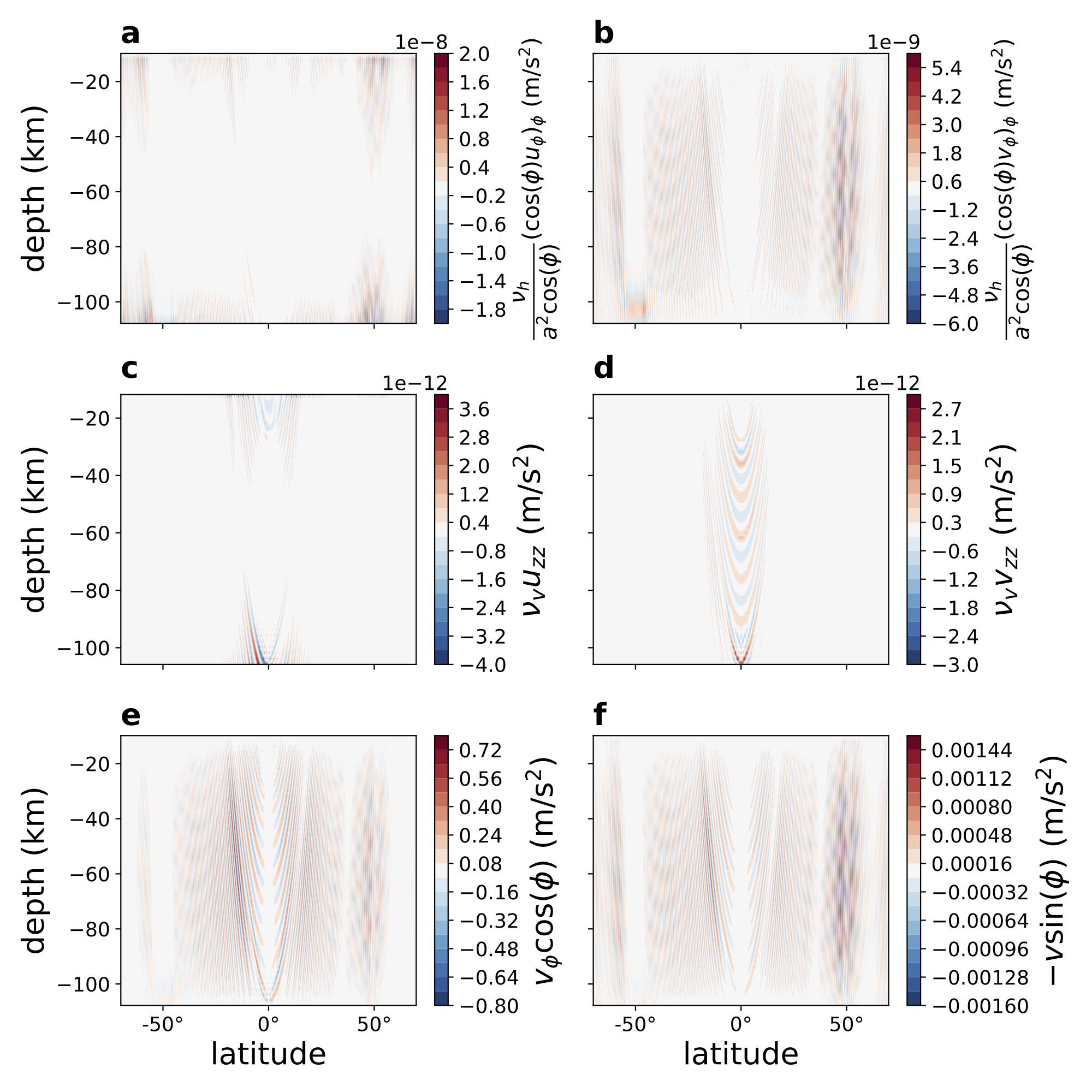}}
\caption{
  {\bf Viscosity terms in the momentum equations and terms in the continuity equation.} Snapshots of the: {\bf a} meridional viscosity of the zonal velocity, $\frac{\nu_h}{a^2\cos(\phi)}(\cos(\phi)u_\phi)_\phi$, {\bf b} meridional viscosity of the meridional velocity, $\frac{\nu_h}{a^2\cos(\phi)}(\cos(\phi)v_\phi)_\phi$, {\bf c} vertical viscosity of the zonal velocity, $\nu_vu_{zz}$, {\bf d} vertical viscosity of the meridional velocity, $\nu_vv_{zz}$. Panels {\bf e} and {\bf f} show terms in the continuity equation and justify our approximation $\frac{1}{a \cos(\phi)} (v \cos(\phi) )_\phi+w_z=0$, that $(v \cos(\phi) )_\phi= v_\phi \cos(\phi)-v\sin(\phi) \approx v_\phi \cos(\phi)$ since $v_\phi \cos(\phi)\gg v\sin(\phi)$. $\phi, z$ are the meridional and vertical coordinates, $u, v, w$ are the zonal, meridional, and vertical velocities, $\nu_h, \nu_v$ are the horizontal and vertical viscosity coefficients, and $a$ is the radius of Europa. Based on Supplement Figs. \ref{fig:fig-2d-terms-dt-advection}--\ref{fig:fig-2d-terms-diff}, 
the most dominant terms in the zonal momentum equation are the Coriolis terms (Supplementary Fig. \ref{fig:fig-2d-terms-coriolis-p}a,b),
  the most dominant terms in the meridional momentum equation are the Coriolis and the pressure gradient terms (Supplementary Fig. \ref{fig:fig-2d-terms-coriolis-p}c,d) which nearly balance each other. The next dominant terms in the zonal and meridional momentum equations are the horizontal viscosity terms (Supplementary Fig. \ref{fig:fig-2d-terms-diff}a,b).
\label{fig:fig-2d-terms-diff}}
\end{figure}

\clearpage
\centerline{\Large{Supplementary References}}


\begin{thebibliography}{10}
\expandafter\ifx\csname url\endcsname\relax
  \def\url#1{\texttt{#1}}\fi
\expandafter\ifx\csname urlprefix\endcsname\relax\def\urlprefix{URL }\fi
\providecommand{\bibinfo}[2]{#2}
\providecommand{\eprint}[2][]{\url{#2}}

\bibitem{Chyba-Phillips-2001:possible}
\bibinfo{author}{Chyba, C.~F.} \& \bibinfo{author}{Phillips, C.~B.}
\newblock \bibinfo{title}{Possible ecosystems and the search for life on
  {Europa}}.
\newblock \emph{\bibinfo{journal}{Proc. Natl. Acad. Sci. U.S.A.}}
  \textbf{\bibinfo{volume}{98}}, \bibinfo{pages}{801--804}
  (\bibinfo{year}{2001}).

\bibitem{Hand-Chyba-Priscu-et-al-2009:astrobiology}
\bibinfo{author}{Hand, K.}, \bibinfo{author}{Chyba, C.},
  \bibinfo{author}{Priscu, J.}, \bibinfo{author}{Carlson, R.} \&
  \bibinfo{author}{Nealson, K.}
\newblock \bibinfo{title}{Astrobiology and the potential for life on {Europa}}.
\newblock \emph{\bibinfo{journal}{Europa. University of Arizona Press, Tucson}}
  \bibinfo{pages}{589--629} (\bibinfo{year}{2009}).

\bibitem{Pappalardo-Vance-Bagenal-et-al-2013:science}
\bibinfo{author}{Pappalardo, R.} \emph{et~al.}
\newblock \bibinfo{title}{Science potential from a {Europa} lander}.
\newblock \emph{\bibinfo{journal}{Astrobiology}} \textbf{\bibinfo{volume}{13}},
  \bibinfo{pages}{740--773} (\bibinfo{year}{2013}).

\bibitem{Cassen-Reynolds-Peale-1979:there}
\bibinfo{author}{Cassen, P.}, \bibinfo{author}{Reynolds, R.~T.} \&
  \bibinfo{author}{Peale, S.}
\newblock \bibinfo{title}{Is there liquid water on {Europa}?}
\newblock \emph{\bibinfo{journal}{Geophys. Res. Lett.}}
  \textbf{\bibinfo{volume}{6}}, \bibinfo{pages}{731--734}
  (\bibinfo{year}{1979}).

\bibitem{Carr-Belton-Chapman-et-al-1998:evidence}
\bibinfo{author}{Carr, M.~H.} \emph{et~al.}
\newblock \bibinfo{title}{Evidence for a subsurface ocean on {Europa}}.
\newblock \emph{\bibinfo{journal}{Nature}} \textbf{\bibinfo{volume}{391}},
  \bibinfo{pages}{363--365} (\bibinfo{year}{1998}).

\bibitem{Kivelson-Khurana-Russell-et-al-2000:galileo}
\bibinfo{author}{Kivelson, M.~G.} \emph{et~al.}
\newblock \bibinfo{title}{Galileo magnetometer measurements: A stronger case
  for a subsurface ocean at {Europa}}.
\newblock \emph{\bibinfo{journal}{Science}} \textbf{\bibinfo{volume}{289}},
  \bibinfo{pages}{1340--1343} (\bibinfo{year}{2000}).

\bibitem{Billings-Kattenhorn-2005:great}
\bibinfo{author}{Billings, S.~E.} \& \bibinfo{author}{Kattenhorn, S.~A.}
\newblock \bibinfo{title}{The great thickness debate: {Ice} shell thickness
  models for {Europa} and comparisons with estimates based on flexure at
  ridges}.
\newblock \emph{\bibinfo{journal}{Icarus}} \textbf{\bibinfo{volume}{177}},
  \bibinfo{pages}{397--412} (\bibinfo{year}{2005}).

\bibitem{Hussmann-Spohn-Wieczerkowski-2002:thermal}
\bibinfo{author}{Hussmann, H.}, \bibinfo{author}{Spohn, T.} \&
  \bibinfo{author}{Wieczerkowski, K.}
\newblock \bibinfo{title}{Thermal equilibrium states of {Europa}'s ice shell:
  Implications for internal ocean thickness and surface heat flow}.
\newblock \emph{\bibinfo{journal}{Icarus}} \textbf{\bibinfo{volume}{156}},
  \bibinfo{pages}{143--151} (\bibinfo{year}{2002}).

\bibitem{Tobie-Choblet-Sotin-2003:tidally}
\bibinfo{author}{Tobie, G.}, \bibinfo{author}{Choblet, G.} \&
  \bibinfo{author}{Sotin, C.}
\newblock \bibinfo{title}{Tidally heated convection: Constraints on {Europa}'s
  ice shell thickness}.
\newblock \emph{\bibinfo{journal}{J. Geophys. Res.}}
  \textbf{\bibinfo{volume}{108}}, \bibinfo{pages}{5124} (\bibinfo{year}{2003}).

\bibitem{Bierhaus-Zahnle-Chapman-et-al-2009:europa}
\bibinfo{author}{Bierhaus, E.~B.} \emph{et~al.}
\newblock \bibinfo{title}{Europa’s crater distributions and surface ages}.
\newblock In \emph{\bibinfo{booktitle}{Europa}}, \bibinfo{pages}{161}
  (\bibinfo{publisher}{University of Arizona Press Tucson},
  \bibinfo{year}{2009}).

\bibitem{Schmidt-Blankenship-Patterson-et-al-2011:active}
\bibinfo{author}{Schmidt, B.}, \bibinfo{author}{Blankenship, D.},
  \bibinfo{author}{Patterson, G.} \& \bibinfo{author}{Schenk, P.}
\newblock \bibinfo{title}{Active formation of 'chaos terrain' over shallow
  subsurface water on {Europa}}.
\newblock \emph{\bibinfo{journal}{Nature}} \textbf{\bibinfo{volume}{479}},
  \bibinfo{pages}{502--505} (\bibinfo{year}{2011}).

\bibitem{Khurana-Kivelson-Stevenson-et-al-1998:induced}
\bibinfo{author}{Khurana, K.} \emph{et~al.}
\newblock \bibinfo{title}{Induced magnetic fields as evidence for subsurface
  oceans in {Europa} and {Callisto}}.
\newblock \emph{\bibinfo{journal}{Nature}} \textbf{\bibinfo{volume}{395}},
  \bibinfo{pages}{777--780} (\bibinfo{year}{1998}).

\bibitem{Pappalardo-Belton-Breneman-et-al-1999:does}
\bibinfo{author}{Pappalardo, R.} \emph{et~al.}
\newblock \bibinfo{title}{Does {Europa} have a subsurface ocean? {Evaluation}
  of the geological evidence}.
\newblock \emph{\bibinfo{journal}{J. Geophys. Res.}}
  \textbf{\bibinfo{volume}{104}}, \bibinfo{pages}{24015--24055}
  (\bibinfo{year}{1999}).

\bibitem{Roth-Saur-Retherford-et-al-2014:transient}
\bibinfo{author}{Roth, L.} \emph{et~al.}
\newblock \bibinfo{title}{Transient water vapor at {Europa's} south pole}.
\newblock \emph{\bibinfo{journal}{Science}} \textbf{\bibinfo{volume}{343}},
  \bibinfo{pages}{171--174} (\bibinfo{year}{2014}).

\bibitem{Sparks-Hand-McGrath-et-al-2016:probing}
\bibinfo{author}{Sparks, W.} \emph{et~al.}
\newblock \bibinfo{title}{Probing for evidence of plumes on {Europa with
  HST/STIS}}.
\newblock \emph{\bibinfo{journal}{The Astrophysical Journal}}
  \textbf{\bibinfo{volume}{829}}, \bibinfo{pages}{121} (\bibinfo{year}{2016}).

\bibitem{Thomson-Delaney-2001:evidence}
\bibinfo{author}{Thomson, R.~E.} \& \bibinfo{author}{Delaney, J.~R.}
\newblock \bibinfo{title}{Evidence for a weakly stratified {Europan} ocean
  sustained by seafloor heat flux}.
\newblock \emph{\bibinfo{journal}{J. Geophys. Res.}}
  \textbf{\bibinfo{volume}{106}}, \bibinfo{pages}{12355--12365}
  (\bibinfo{year}{2001}).

\bibitem{Goodman-Collins-Marshall-et-al-2004:hydrothermal}
\bibinfo{author}{Goodman, J.~C.}, \bibinfo{author}{Collins, G.~C.},
  \bibinfo{author}{Marshall, J.} \& \bibinfo{author}{Pierrehumbert, R.~T.}
\newblock \bibinfo{title}{Hydrothermal plume dynamics on {Europa}:
  {Implications} for chaos formation}.
\newblock \emph{\bibinfo{journal}{J. Geophys. Res.}}
  \textbf{\bibinfo{volume}{109}} (\bibinfo{year}{2004}).

\bibitem{Melosh-Ekholm-Showman-et-al-2004:temperature}
\bibinfo{author}{Melosh, H.}, \bibinfo{author}{Ekholm, A.},
  \bibinfo{author}{Showman, A.} \& \bibinfo{author}{Lorenz, R.}
\newblock \bibinfo{title}{The temperature of {Europa's} subsurface water
  ocean}.
\newblock \emph{\bibinfo{journal}{Icarus}} \textbf{\bibinfo{volume}{168}},
  \bibinfo{pages}{498--502} (\bibinfo{year}{2004}).

\bibitem{Tyler-2008:strong}
\bibinfo{author}{Tyler, R.~H.}
\newblock \bibinfo{title}{Strong ocean tidal flow and heating on moons of the
  outer planets}.
\newblock \emph{\bibinfo{journal}{Nature}} \textbf{\bibinfo{volume}{456}},
  \bibinfo{pages}{770--772} (\bibinfo{year}{2008}).

\bibitem{Vance-Goodman-2009:oceanography}
\bibinfo{author}{Vance, S.} \& \bibinfo{author}{Goodman, J.}
\newblock \bibinfo{title}{Oceanography of an ice-covered moon}.
\newblock In \bibinfo{editor}{Pappalardo, R.~T.}, \bibinfo{editor}{McKinnon,
  W.~B.} \& \bibinfo{editor}{Khurana, K.} (eds.)
  \emph{\bibinfo{booktitle}{Europa}}, \bibinfo{pages}{459--482}
  (\bibinfo{publisher}{The University of Arizona Press, Tucson, AZ},
  \bibinfo{year}{2009}).

\bibitem{Goodman-2012:tilted}
\bibinfo{author}{Goodman, J.~C.}
\newblock \bibinfo{title}{Tilted geostrophic convection in icy world oceans
  caused by the horizontal component of the planetary rotation vector}.
\newblock In \emph{\bibinfo{booktitle}{American Geophysical Union, Fall Meeting
  2012, abstract}}, \bibinfo{pages}{P51A--2017} (\bibinfo{year}{2012}).

\bibitem{Goodman-Lenferink-2012:numerical}
\bibinfo{author}{Goodman, J.~C.} \& \bibinfo{author}{Lenferink, E.}
\newblock \bibinfo{title}{Numerical simulations of marine hydrothermal plumes
  for {Europa} and other icy worlds}.
\newblock \emph{\bibinfo{journal}{Icarus}} \textbf{\bibinfo{volume}{221}},
  \bibinfo{pages}{970--983} (\bibinfo{year}{2012}).

\bibitem{Soderlund-Schmidt-Wicht-et-al-2014:ocean}
\bibinfo{author}{Soderlund, K.~M.}, \bibinfo{author}{Schmidt, B.~E.},
  \bibinfo{author}{Wicht, J.} \& \bibinfo{author}{Blankenship, D.~D.}
\newblock \bibinfo{title}{Ocean-driven heating of {Europa's} icy shell at low
  latitudes}.
\newblock \emph{\bibinfo{journal}{Nature Geoscience}}
  \textbf{\bibinfo{volume}{7}}, \bibinfo{pages}{16--19} (\bibinfo{year}{2014}).

\bibitem{Gissinger-Petitdemange-2019:magnetically}
\bibinfo{author}{Gissinger, C.} \& \bibinfo{author}{Petitdemange, L.}
\newblock \bibinfo{title}{A magnetically driven equatorial jet in europa’s
  ocean}.
\newblock \emph{\bibinfo{journal}{Nature Astronomy}}
  \textbf{\bibinfo{volume}{3}}, \bibinfo{pages}{401} (\bibinfo{year}{2019}).

\bibitem{Soderlund-2019:ocean}
\bibinfo{author}{Soderlund, K.~M.}
\newblock \bibinfo{title}{Ocean dynamics of outer solar system satellites}.
\newblock \emph{\bibinfo{journal}{Geophys. Res. Lett.}}
  \bibinfo{pages}{doi:10.1029/2018GL081880} (\bibinfo{year}{2019}).

\bibitem{Rovira-Navarro-Rieutord-Gerkema-et-al-2019:do}
\bibinfo{author}{Rovira-Navarro, M.} \emph{et~al.}
\newblock \bibinfo{title}{Do tidally-generated inertial waves heat the
  subsurface oceans of europa and enceladus?}
\newblock \emph{\bibinfo{journal}{Icarus}} \textbf{\bibinfo{volume}{321}},
  \bibinfo{pages}{126--140} (\bibinfo{year}{2019}).

\bibitem{Lemasquerier-Grannan-Vidal-et-al-2017:libration}
\bibinfo{author}{Lemasquerier, D.} \emph{et~al.}
\newblock \bibinfo{title}{Libration-driven flows in ellipsoidal shells}.
\newblock \emph{\bibinfo{journal}{Journal of Geophysical Research: Planets}}
  \textbf{\bibinfo{volume}{122}}, \bibinfo{pages}{1926--1950}
  (\bibinfo{year}{2017}).

\bibitem{Marshall-Adcroft-Hill-et-al-1997:finite}
\bibinfo{author}{Marshall, J.}, \bibinfo{author}{Adcroft, A.},
  \bibinfo{author}{Hill, C.}, \bibinfo{author}{Perelman, L.} \&
  \bibinfo{author}{Heisey, C.}
\newblock \bibinfo{title}{A finite-volume, incompressible {Navier} {Stokes}
  model for studies of the ocean on parallel computers}.
\newblock \emph{\bibinfo{journal}{J. Geophys. Res.}}
  \textbf{\bibinfo{volume}{102, C3}}, \bibinfo{pages}{5,753--5,766}
  (\bibinfo{year}{1997}).

\bibitem{MITgcm-manual-github:mitgcm}
\bibinfo{author}{{MITgcm~Group}}.
\newblock \bibinfo{title}{{MITgcm} {U}ser {M}anual}.
\newblock \bibinfo{type}{Online documentation},
  \bibinfo{institution}{{MIT}/{EAPS}}, \bibinfo{address}{Cambridge, MA 02139,
  USA} (\bibinfo{year}{2021}).
\newblock \bibinfo{note}{\rm{https://mitgcm.readthedocs.io/en/latest/}}.

\bibitem{Losch-2008:modeling}
\bibinfo{author}{Losch, M.}
\newblock \bibinfo{title}{Modeling ice shelf cavities in a z-coordinate ocean
  general circulation model}.
\newblock \emph{\bibinfo{journal}{J. Geophys. Res.}}
  \textbf{\bibinfo{volume}{113}}, \bibinfo{pages}{C08043}
  (\bibinfo{year}{2008}).

\bibitem{Ashkenazy-Sayag-Tziperman-2018:dynamics}
\bibinfo{author}{Ashkenazy, Y.}, \bibinfo{author}{Sayag, R.} \&
  \bibinfo{author}{Tziperman, E.}
\newblock \bibinfo{title}{Dynamics of the global meridional ice flow of
  {Europa}'s icy shell}.
\newblock \emph{\bibinfo{journal}{Nature Astronomy}}
  \textbf{\bibinfo{volume}{2}}, \bibinfo{pages}{43} (\bibinfo{year}{2018}).

\bibitem{Hand-Chyba-2007:empirical}
\bibinfo{author}{Hand, K.~P.} \& \bibinfo{author}{Chyba, C.~F.}
\newblock \bibinfo{title}{Empirical constraints on the salinity of the europan
  ocean and implications for a thin ice shell}.
\newblock \emph{\bibinfo{journal}{Icarus}} \textbf{\bibinfo{volume}{189}},
  \bibinfo{pages}{424--438} (\bibinfo{year}{2007}).

\bibitem{Ojakangas-Stevenson-1989:thermal}
\bibinfo{author}{Ojakangas, G.~W.} \& \bibinfo{author}{Stevenson, D.~J.}
\newblock \bibinfo{title}{Thermal state of an ice shell on {Europa}}.
\newblock \emph{\bibinfo{journal}{Icarus}} \textbf{\bibinfo{volume}{81}},
  \bibinfo{pages}{220--241} (\bibinfo{year}{1989}).

\bibitem{Ashkenazy-2019:surface}
\bibinfo{author}{Ashkenazy, Y.}
\newblock \bibinfo{title}{The surface temperature of {Europa}}.
\newblock \emph{\bibinfo{journal}{Heliyon}} \textbf{\bibinfo{volume}{5}},
  \bibinfo{pages}{arxiv.org/abs/1608.07372} (\bibinfo{year}{2018}).

\bibitem{Jansen-2016:turbulent}
\bibinfo{author}{Jansen, M.~F.}
\newblock \bibinfo{title}{The turbulent circulation of a {Snowball} {Earth}
  ocean}.
\newblock \emph{\bibinfo{journal}{J. Phys. Oceanogr.}}
  \textbf{\bibinfo{volume}{46}}, \bibinfo{pages}{1917--1933}
  (\bibinfo{year}{2016}).

\bibitem{Ashkenazy-Tziperman-2016:variability}
\bibinfo{author}{Ashkenazy, Y.} \& \bibinfo{author}{Tziperman, E.}
\newblock \bibinfo{title}{Variability, instabilities and eddies in a {Snowball}
  ocean}.
\newblock \emph{\bibinfo{journal}{J. Climate}} \textbf{\bibinfo{volume}{29}},
  \bibinfo{pages}{869--888, doi: http://dx.doi.org/10.1175/JCLI--D--15--0308.1}
  (\bibinfo{year}{2016}).

\bibitem{Vance-Brown-2005:layering}
\bibinfo{author}{Vance, S.} \& \bibinfo{author}{Brown, J.~M.}
\newblock \bibinfo{title}{Layering and double-diffusion style convection in
  {Europa}'s ocean}.
\newblock \emph{\bibinfo{journal}{Icarus}} \textbf{\bibinfo{volume}{177}},
  \bibinfo{pages}{506--514} (\bibinfo{year}{2005}).

\bibitem{Vonstorch-Eden-Fast-et-al-2012:estimate}
\bibinfo{author}{vonStorch, J.~S.} \emph{et~al.}
\newblock \bibinfo{title}{An estimate of the lorenz energy cycle for the world
  ocean based on the storm/ncep simulation}.
\newblock \emph{\bibinfo{journal}{J. Phys. Oceanogr.}}
  \textbf{\bibinfo{volume}{42}}, \bibinfo{pages}{2185--2205}
  (\bibinfo{year}{2012}).

\bibitem{Huang-1999:mixing}
\bibinfo{author}{Huang, R.~X.}
\newblock \bibinfo{title}{Mixing and energetics of the oceanic thermohaline
  circulation}.
\newblock \emph{\bibinfo{journal}{J. Phys. Oceanogr.}}
  \textbf{\bibinfo{volume}{29}}, \bibinfo{pages}{727--746}
  (\bibinfo{year}{1999}).

\bibitem{Sotin-Tobie-Wahr-et-al-2009:tides}
\bibinfo{author}{Sotin, C.}, \bibinfo{author}{Tobie, G.},
  \bibinfo{author}{Wahr, J.} \& \bibinfo{author}{McKinnon, W.~B.}
\newblock \bibinfo{title}{Tides and tidal heating on {Europa}}.
\newblock In \bibinfo{editor}{Pappalardo, R.~T.}, \bibinfo{editor}{McKinnon,
  W.~B.} \& \bibinfo{editor}{Khurana, K.} (eds.)
  \emph{\bibinfo{booktitle}{Europa}}, \bibinfo{pages}{85--117}
  (\bibinfo{publisher}{The University of Arizona Press, Tucson, AZ},
  \bibinfo{year}{2009}).

\bibitem{Chen-Nimmo-Glatzmaier-2014:tidal}
\bibinfo{author}{Chen, E. M.~A.}, \bibinfo{author}{Nimmo, F.} \&
  \bibinfo{author}{Glatzmaier, G.~A.}
\newblock \bibinfo{title}{Tidal heating in icy satellite oceans}.
\newblock \emph{\bibinfo{journal}{Icarus}} \textbf{\bibinfo{volume}{229}},
  \bibinfo{pages}{11--30} (\bibinfo{year}{2014}).

\bibitem{Ashkenazy-Gildor-Losch-et-al-2013:dynamics}
\bibinfo{author}{Ashkenazy, Y.} \emph{et~al.}
\newblock \bibinfo{title}{Dynamics of a {Snowball Earth} ocean}.
\newblock \emph{\bibinfo{journal}{Nature}} \textbf{\bibinfo{volume}{495}},
  \bibinfo{pages}{90--93, doi:10.1038/nature11894} (\bibinfo{year}{2013}).

\bibitem{Kang-Mittal-Bire-et-al-2021:how}
\bibinfo{author}{Kang, W.}, \bibinfo{author}{Mittal, T.},
  \bibinfo{author}{Bire, S.}, \bibinfo{author}{Michel, J.} \&
  \bibinfo{author}{Marshall, J.}
\newblock \bibinfo{title}{How does salinity shape ocean circulation and ice
  geometry on enceladus and other icy satellites?}
\newblock \emph{\bibinfo{journal}{arXiv:2104.07008 [astro-ph.EP]}}
  (\bibinfo{year}{2021}).

\bibitem{Zeng-Jansen-2021:ocean}
\bibinfo{author}{Zeng, Y.} \& \bibinfo{author}{Jansen, M.~F.}
\newblock \bibinfo{title}{Ocean circulation on enceladus with a high versus low
  salinity ocean}.
\newblock \emph{\bibinfo{journal}{arXiv:2101.10530 [astro-ph.EP]}}
  (\bibinfo{year}{2021}).

\bibitem{Lobo-Thompson-Vance-et-al-2020:pole}
\bibinfo{author}{Lobo, A.~H.}, \bibinfo{author}{Thompson, A.~F.},
  \bibinfo{author}{Vance, S.~D.} \& \bibinfo{author}{Tharimena, S.}
\newblock \bibinfo{title}{A pole-to-equator ocean overturning circulation on
  enceladus} (\bibinfo{year}{2020}).
\newblock \eprint{2007.06173}.

\bibitem{Pappalardo-Prockter-Senske-et-al-2016:science}
\bibinfo{author}{Pappalardo, R.} \emph{et~al.}
\newblock \bibinfo{title}{Science objectives and capabilities of the {NASA}
  {Europa} mission}.
\newblock In \emph{\bibinfo{booktitle}{Lunar and Planetary Science
  Conference}}, vol.~\bibinfo{volume}{47}, \bibinfo{pages}{3058}
  (\bibinfo{year}{2016}).

\bibitem{Howell-Pappalardo-2020:nasas}
\bibinfo{author}{Howell, S.~M.} \& \bibinfo{author}{Pappalardo, R.~T.}
\newblock \bibinfo{title}{Nasa's europa clipper—a mission to a potentially
  habitable ocean world}.
\newblock \emph{\bibinfo{journal}{Nature communications}}
  \textbf{\bibinfo{volume}{11}}, \bibinfo{pages}{1--4} (\bibinfo{year}{2020}).

\bibitem{Grasset-Dougherty-Coustenis-et-al-2013:jupiter}
\bibinfo{author}{Grasset, O.} \emph{et~al.}
\newblock \bibinfo{title}{{JUpiter} {ICy} moons {Explorer} ({JUICE}): {An}
  {ESA} mission to orbit {Ganymede} and to characterise the {Jupiter} system}.
\newblock \emph{\bibinfo{journal}{Planetary and Space Science}}
  \textbf{\bibinfo{volume}{78}}, \bibinfo{pages}{1--21} (\bibinfo{year}{2013}).

\bibitem{Vance-Styczinski-Bills-et-al-2021:magnetic}
\bibinfo{author}{Vance, S.~D.} \emph{et~al.}
\newblock \bibinfo{title}{Magnetic induction responses of {Jupiter's} ocean
  moons including effects from adiabatic convection}.
\newblock \emph{\bibinfo{journal}{Journal of Geophysical Research: Planets}}
  \textbf{\bibinfo{volume}{126}}, \bibinfo{pages}{e2020JE006418}
  (\bibinfo{year}{2021}).

\bibitem{Kaspi-2008:turbulent}
\bibinfo{author}{Kaspi, Y.}
\newblock \emph{\bibinfo{title}{Turbulent convection in an anelastic rotating
  sphere: A model for the circulation on the giant planets}}.
\newblock Ph.D. thesis, \bibinfo{school}{Massachusetts Institute of Technology
  and Woods Hole Oceanographic Institution} (\bibinfo{year}{2008}).

\bibitem{Kaspi-Flierl-Showman-2009:deep}
\bibinfo{author}{Kaspi, Y.}, \bibinfo{author}{Flierl, G.~R.} \&
  \bibinfo{author}{Showman, A.~P.}
\newblock \bibinfo{title}{The deep wind structure of the giant planets: Results
  from an anelastic general circulation model}.
\newblock \emph{\bibinfo{journal}{Icarus}} \textbf{\bibinfo{volume}{202}},
  \bibinfo{pages}{525--542} (\bibinfo{year}{2009}).

\bibitem{Zalucha-Gulbis-2012:comparison}
\bibinfo{author}{Zalucha, A.} \& \bibinfo{author}{Gulbis, A.}
\newblock \bibinfo{title}{Comparison of a simple 2-{D} {Pluto} general
  circulation model with stellar occultation light curves and implications for
  atmospheric circulation}.
\newblock \emph{\bibinfo{journal}{J. Geophys. Res.}}
  \textbf{\bibinfo{volume}{117}} (\bibinfo{year}{2012}).

\bibitem{Zalucha-Michaels-2013:3d}
\bibinfo{author}{Zalucha, A.~M.} \& \bibinfo{author}{Michaels, T.~I.}
\newblock \bibinfo{title}{A {3D} general circulation model for {Pluto} and
  {Triton} with fixed volatile abundance and simplified surface forcing}.
\newblock \emph{\bibinfo{journal}{Icarus}} \textbf{\bibinfo{volume}{223}},
  \bibinfo{pages}{819--831} (\bibinfo{year}{2013}).

\bibitem{Showman-Fortney-Lian-et-al-2009:atmospheric}
\bibinfo{author}{Showman, A.~P.} \emph{et~al.}
\newblock \bibinfo{title}{Atmospheric circulation of hot {Jupiters}: {Coupled}
  radiative-dynamical general circulation model simulations of {HD} 189733b and
  {HD} 209458b}.
\newblock \emph{\bibinfo{journal}{The Astrophysical Journal}}
  \textbf{\bibinfo{volume}{699}}, \bibinfo{pages}{564} (\bibinfo{year}{2009}).

\bibitem{Parmentier-Fortney-Showman-et-al-2016:transitions}
\bibinfo{author}{Parmentier, V.}, \bibinfo{author}{Fortney, J.~J.},
  \bibinfo{author}{Showman, A.~P.}, \bibinfo{author}{Morley, C.} \&
  \bibinfo{author}{Marley, M.~S.}
\newblock \bibinfo{title}{Transitions in the cloud composition of hot
  {Jupiters}}.
\newblock \emph{\bibinfo{journal}{The Astrophysical Journal}}
  \textbf{\bibinfo{volume}{828}}, \bibinfo{pages}{22} (\bibinfo{year}{2016}).

\bibitem{Ashkenazy-Gildor-Losch-et-al-2014:ocean}
\bibinfo{author}{Ashkenazy, Y.}, \bibinfo{author}{Gildor, H.},
  \bibinfo{author}{Losch, M.} \& \bibinfo{author}{Tziperman, E.}
\newblock \bibinfo{title}{Ocean circulation under globally glaciated {Snowball
  Earth} conditions: steady state solutions}.
\newblock \emph{\bibinfo{journal}{J. Phys. Oceanogr.}}
  \textbf{\bibinfo{volume}{44}}, \bibinfo{pages}{24--43}
  (\bibinfo{year}{2014}).

\bibitem{Large-Mcwilliams-Doney-1994:oceanic}
\bibinfo{author}{Large, W.~G.}, \bibinfo{author}{Mcwilliams, J.~C.} \&
  \bibinfo{author}{Doney, S.~C.}
\newblock \bibinfo{title}{Oceanic vertical mixing: A review and a model with a
  nonlocal boundary-layer parameterization}.
\newblock \emph{\bibinfo{journal}{Rev. Geophys.}}
  \textbf{\bibinfo{volume}{32}}, \bibinfo{pages}{363--403}
  (\bibinfo{year}{1994}).

\bibitem{Gent-Mcwilliams-1990:isopycnal}
\bibinfo{author}{Gent, P.~R.} \& \bibinfo{author}{{McWilliams}, J.~C.}
\newblock \bibinfo{title}{Isopycnal mixing in ocean circulation models}.
\newblock \emph{\bibinfo{journal}{J. Phys. Oceanogr.}}
  \textbf{\bibinfo{volume}{20}}, \bibinfo{pages}{150--155}
  (\bibinfo{year}{1990}).

\bibitem{Greenberg-Geissler-Hoppa-et-al-2002:tidal}
\bibinfo{author}{Greenberg, R.}, \bibinfo{author}{Geissler, P.},
  \bibinfo{author}{Hoppa, G.} \& \bibinfo{author}{Tufts, B.}
\newblock \bibinfo{title}{Tidal-tectonic processes and their implications for
  the character of {Europa}'s icy crust}.
\newblock \emph{\bibinfo{journal}{Reviews of Geophysics}}
  \textbf{\bibinfo{volume}{40}} (\bibinfo{year}{2002}).

\bibitem{Barr-Showman-2009:heat}
\bibinfo{author}{Barr, A.~C.} \& \bibinfo{author}{Showman, A.~P.}
\newblock \bibinfo{title}{Heat transfer in {Europa's} icy shell}.
\newblock In \bibinfo{editor}{Pappalardo, R.~T.}, \bibinfo{editor}{McKinnon,
  W.~B.} \& \bibinfo{editor}{Khurana, K.} (eds.)
  \emph{\bibinfo{booktitle}{Europa}}, \bibinfo{pages}{405--430}
  (\bibinfo{publisher}{The University of Arizona Press, Tucson, AZ},
  \bibinfo{year}{2009}).

\bibitem{Christensen-Wicht-2015:numerical}
\bibinfo{author}{Christensen, U.~R.} \& \bibinfo{author}{Wicht, J.}
\newblock \bibinfo{title}{Numerical dynamo simulations}.
\newblock \emph{\bibinfo{journal}{Treatise on Geophysics (Second Edition)}}
  \textbf{\bibinfo{volume}{8}}, \bibinfo{pages}{245--277}
  (\bibinfo{year}{2015}).

\bibitem{Jackett-Mcdougall-1995:minimal}
\bibinfo{author}{Jackett, D.~R.} \& \bibinfo{author}{McDougall, T.~J.}
\newblock \bibinfo{title}{Minimal adjustment of hydrographic profiles to
  achieve static stability}.
\newblock \emph{\bibinfo{journal}{J. Atmos. Ocean Tech.}}
  \textbf{\bibinfo{volume}{12}}, \bibinfo{pages}{381--389}
  (\bibinfo{year}{1995}).

\bibitem{Zolotov-Kargel-2009:chemical}
\bibinfo{author}{Zolotov, M.~Y.} \& \bibinfo{author}{Kargel, J.}
\newblock \bibinfo{title}{On the chemical composition of {Europa}'s icy shell,
  ocean, and underlying rocks}.
\newblock \emph{\bibinfo{journal}{Europa}} \bibinfo{pages}{431--457}
  (\bibinfo{year}{2009}).

\bibitem{Zhang-Schubert-2000:magnetohydrodynamics}
\bibinfo{author}{Zhang, K.} \& \bibinfo{author}{Schubert, G.}
\newblock \bibinfo{title}{Magnetohydrodynamics in rapidly rotating spherical
  systems}.
\newblock \emph{\bibinfo{journal}{Annual Review of Fluid Mechanics}}
  \textbf{\bibinfo{volume}{32}}, \bibinfo{pages}{409--443}
  (\bibinfo{year}{2000}).

\bibitem{Hinze-1975:turbulence}
\bibinfo{author}{Hinze, J.~O.}
\newblock \emph{\bibinfo{title}{Turbulence}} (\bibinfo{publisher}{McGrow-Hill},
  \bibinfo{year}{1975}).

\bibitem{Lemmin-1989:dynamics}
\bibinfo{author}{Lemmin, U.}
\newblock \bibinfo{title}{Dynamics of horizontal turbulent mixing in a
  nearshore zone of {Lake} {Geneva}}.
\newblock \emph{\bibinfo{journal}{Limnol. Oceanog.}}
  \textbf{\bibinfo{volume}{34}}, \bibinfo{pages}{420--434}
  (\bibinfo{year}{1989}).

\bibitem{Abernathey-Marshall-2013:global}
\bibinfo{author}{Abernathey, R.~P.} \& \bibinfo{author}{Marshall, J.}
\newblock \bibinfo{title}{Global surface eddy diffusivities derived from
  satellite altimetry}.
\newblock \emph{\bibinfo{journal}{Journal of Geophysical Research: Oceans}}
  \textbf{\bibinfo{volume}{118}}, \bibinfo{pages}{901--916}
  (\bibinfo{year}{2013}).

\bibitem{Christensen-Aubert-2006:scaling}
\bibinfo{author}{Christensen, U.~R.} \& \bibinfo{author}{Aubert, J.}
\newblock \bibinfo{title}{Scaling properties of convection-driven dynamos in
  rotating spherical shells and application to planetary magnetic fields}.
\newblock \emph{\bibinfo{journal}{Geophysical Journal International}}
  \textbf{\bibinfo{volume}{166}}, \bibinfo{pages}{97--114}
  (\bibinfo{year}{2006}).

\end{thebibliography}

\begin{thebibliography}{1}
\expandafter\ifx\csname url\endcsname\relax
  \def\url#1{\texttt{#1}}\fi
\expandafter\ifx\csname urlprefix\endcsname\relax\def\urlprefix{URL }\fi
\providecommand{\bibinfo}[2]{#2}
\providecommand{\eprint}[2][]{\url{#2}}

\bibitem{Vance-Goodman-2009:oceanography}
\bibinfo{author}{Vance, S.} \& \bibinfo{author}{Goodman, J.}
\newblock \bibinfo{title}{Oceanography of an ice-covered moon}.
\newblock In \bibinfo{editor}{Pappalardo, R.~T.}, \bibinfo{editor}{McKinnon,
  W.~B.} \& \bibinfo{editor}{Khurana, K.} (eds.)
  \emph{\bibinfo{booktitle}{Europa}}, \bibinfo{pages}{459--482}
  (\bibinfo{publisher}{The University of Arizona Press, Tucson, AZ},
  \bibinfo{year}{2009}).

\end{thebibliography}

\end{document}